
Manuscript is plain TeX with macros.  Figures start
at FIGURES BEGIN HERE, with instructions on what to do to get them.

%
%
  \font\twelverm=cmr10 scaled 1200       \font\twelvei=cmmi10 scaled 1200
  \font\twelvesy=cmsy10 scaled 1200      \font\twelveex=cmex10 scaled 1200
  \font\twelvebf=cmbx10 scaled 1200      \font\twelvesl=cmsl10 scaled 1200
  \font\twelvett=cmtt10 scaled 1200      \font\twelveit=cmti10 scaled 1200
  \font\twelvemib=cmmib10 scaled 1200
  \font\elevenmib=cmmib10 scaled 1095
  \font\tenmib=cmmib10
  \font\eightmib=cmmib10 scaled 800
  
\font\elevenrm=cmr10 scaled 1095    \font\eleveni=cmmi10 scaled 1095
\font\elevensy=cmsy10 scaled 1095

%
%

\font\seventeeni=cmmi10 scaled \magstep3

\font\seventeensy=cmsy10 scaled \magstep3

\font\seventeenmib=cmmib10 scaled \magstep3

\newfam\cpfam%



\skewchar\eleveni='177   \skewchar\elevensy='60
\skewchar\elevenmib='177  \skewchar\seventeensy='60
\skewchar\seventeenmib='177
\skewchar\seventeeni='177
     
\newfam\mibfam%
     

  \skewchar\twelvei='177   \skewchar\twelvesy='60
  \skewchar\twelvemib='177
%
%
\def\twelvepoint{\normalbaselineskip=12.4pt
  \abovedisplayskip 12.4pt plus 3pt minus 9pt
  \belowdisplayskip 12.4pt plus 3pt minus 9pt
  \abovedisplayshortskip 0pt plus 3pt
  \belowdisplayshortskip 7.2pt plus 3pt minus 4pt
  \smallskipamount=3.6pt plus 1.2pt minus 1.2pt
  \medskipamount=7.2pt plus 2.4pt minus 2.4pt
  \bigskipamount=14.4pt plus 4.8pt minus 4.8pt
  \def\rm{\fam0\twelverm}          \def\it{\fam\itfam\twelveit}%
  \def\sl{\fam\slfam\twelvesl}     \def\bf{\fam\bffam\twelvebf}%
  \def\mit{\fam 1}                 \def\cal{\fam 2}%
  \def\tt{\twelvett}%
  \def\mib{\fam\mibfam\twelvemib}%
     
  \textfont0=\twelverm   \scriptfont0=\tenrm     \scriptscriptfont0=\sevenrm
  \textfont1=\twelvei    \scriptfont1=\teni      \scriptscriptfont1=\seveni
  \textfont2=\twelvesy   \scriptfont2=\tensy     \scriptscriptfont2=\sevensy
  \textfont3=\twelveex   \scriptfont3=\twelveex  \scriptscriptfont3=\twelveex
  \textfont\itfam=\twelveit
  \textfont\slfam=\twelvesl
  \textfont\bffam=\twelvebf
  \textfont\mibfam=\twelvemib       \scriptfont\mibfam=\tenmib
                                             \scriptscriptfont\mibfam=\eightmib

  \def\xrm{\textfont0=\twelverm\scriptfont0=\tenrm
      \scriptscriptfont0=\sevenrm\rm}
\normalbaselines\rm}
     

\mathchardef\alpha="710B
\mathchardef\beta="710C
\mathchardef\gamma="710D
\mathchardef\delta="710E
\mathchardef\epsilon="710F
\mathchardef\zeta="7110
\mathchardef\eta="7111
\mathchardef\theta="7112
\mathchardef\kappa="7114
\mathchardef\lambda="7115
\mathchardef\mu="7116
\mathchardef\nu="7117
\mathchardef\xi="7118
\mathchardef\pi="7119
\mathchardef\rho="711A
\mathchardef\sigma="711B
\mathchardef\tau="711C
\mathchardef\phi="711E
\mathchardef\chi="711F
\mathchardef\psi="7120
\mathchardef\omega="7121
\mathchardef\varepsilon="7122
\mathchardef\vartheta="7123
\mathchardef\varrho="7125
\mathchardef\varphi="7127
     
\def\physgreek{
\mathchardef\Gamma="7100
\mathchardef\Delta="7101
\mathchardef\Theta="7102
\mathchardef\Lambda="7103
\mathchardef\Xi="7104
\mathchardef\Pi="7105
\mathchardef\Sigma="7106
\mathchardef\Upsilon="7107
\mathchardef\Phi="7108
\mathchardef\Psi="7109
\mathchardef\Omega="710A}
     
     
\def\beginlinemode{\endmode
  \begingroup\parskip=0pt \obeylines\def\\{\par}\def\endmode{\par\endgroup}}
\def\beginparmode{\endmode
  \begingroup \def\endmode{\par\endgroup}}
\let\endmode=\par
{\obeylines\gdef\
{}}
\def\singlespace{\baselineskip=\normalbaselineskip}

\def\oneandahalfspace{\baselineskip=\normalbaselineskip
  \multiply\baselineskip by 3 \divide\baselineskip by 2}
\def\doublespace{\baselineskip=\normalbaselineskip \multiply\baselineskip by 2}

\nopagenumbers
\newcount\firstpageno
\firstpageno=2
\headline={\ifnum\pageno<\firstpageno{\hfil}\else{\hfil\elevenrm\folio\hfil}\fi}
\let\rawfootnote=\footnote             
\def\footnote#1#2{{\singlespace\parindent=0pt
\rawfootnote{#1}{#2}}}
\def\raggedcenter{\leftskip=4em plus 12em \rightskip=\leftskip
  \parindent=0pt \parfillskip=0pt \spaceskip=.3333em \xspaceskip=.5em
  \pretolerance=9999 \tolerance=9999
  \hyphenpenalty=9999 \exhyphenpenalty=9999 }
\def\dateline{\rightline{\ifcase\month\or
  January\or February\or March\or April\or May\or June\or
  July\or August\or September\or October\or November\or December\fi
  \space\number\year}}
\def\received{\vskip 3pt plus 0.2fill
 \centerline{\sl (Received\space\ifcase\month\or
  January\or February\or March\or April\or May\or June\or
  July\or August\or September\or October\or November\or December\fi
  \qquad, \number\year)}}
     
     
\hsize=6.5truein
\hoffset=0.0truein
\vsize=8.9truein
\voffset=0truein
\hfuzz=0.1pt
\vfuzz=0.1pt
\parskip=\medskipamount
\overfullrule=0pt      
     
     
     
\def\title                     
  {\null\vskip 3pt plus 0.1fill
   \beginlinemode \doublespace \raggedcenter \bf}
     
\def\author                    
  {\vskip 6pt plus 0.2fill \beginlinemode
   \singlespace \raggedcenter}
     
\def\affil        
  {\vskip 6pt plus 0.1fill \beginlinemode
   \oneandahalfspace \raggedcenter \it}
     
\def\abstract                  
  {\vskip 6pt plus 0.3fill \beginparmode
   \doublespace \narrower }
     
\def\summary                   
  {\vskip 3pt plus 0.3fill \beginparmode
   \doublespace \narrower SUMMARY: }
     
\def\pacs#1
  {\vskip 3pt plus 0.2fill PACS numbers: #1}
     
\def\endtitlepage              
  {\endpage                    
   \body}
     
\def\body                      
  {\beginparmode}              
     
\def\head#1{                   
  \filbreak\vskip 0.5truein    
  {\immediate\write16{#1}
   \raggedcenter \uppercase{#1}\par}
   \nobreak\vskip 0.25truein\nobreak}

%
%

%
\def\inlinerefs{
  \gdef\refto##1{ [##1]}                
\gdef\refis##1{\indent\hbox to 0pt{\hss##1.~}} 
\gdef\journal##1, ##2, ##3, 1##4##5##6{ 
    {\sl ##1~}{\bf ##2}, ##3 (1##4##5##6)}}    
\def\keywords#1
  {\vskip 3pt plus 0.2fill Keywords: #1}
\gdef\figis#1{\indent\hbox to 0pt{\hss#1.~}} 

\def\figurecaptions     
  {\head{Figure Captions}    
   \beginparmode
   \interlinepenalty=10000
   \frenchspacing \parindent=0pt \leftskip=1truecm
   \parskip=8pt plus 3pt \everypar{\hangindent=\parindent}}
     
%
%
\def\refto#1{$^{#1}$}          
     
\def\references       
  {\head{References}           
   \beginparmode
   \frenchspacing \parindent=0pt \leftskip=1truecm
   \interlinepenalty=10000
   \parskip=8pt plus 3pt \everypar{\hangindent=\parindent}}
     
\gdef\refis#1{\indent\hbox to 0pt{\hss#1.~}} 

\gdef\journal#1, #2, #3, 1#4#5#6{              
    {\sl #1~}{\bf #2}, #3 (1#4#5#6)}          
     
\def\refstylenp{               
  \gdef\refto##1{ [##1]}                               
  \gdef\refis##1{\indent\hbox to 0pt{\hss##1)~}}      
  \gdef\journal##1, ##2, ##3, ##4 {                    
     {\sl ##1~}{\bf ##2~}(##3) ##4 }}
     
\def\refstyleprnp{             
  \gdef\refto##1{ [##1]}                               
  \gdef\refis##1{\indent\hbox to 0pt{\hss##1)~}}      
  \gdef\journal##1, ##2, ##3, 1##4##5##6{              
    {\sl ##1~}{\bf ##2~}(1##4##5##6) ##3}}
     
\def\pr{\journal Phys. Rev., }
     
\def\pra{\journal Phys. Rev. A, }
     
\def\prb{\journal Phys. Rev. B, }

\def\prl{\journal Phys. Rev. Lett., }

\def\revmp{\journal Rev. Mod. Phys., }

\def\prsl{\journal Proc. Roy. Soc. London, }
     
\def\jpsj{\journal J. Phys. Soc. Japan., }

\def\jlowt{\journal J. Low Temp. Phys., }

\def\endreferences{\body}
     
%
%
     
\def\endfigurecaptions{\body}
     
\def\endpage                   
  {\vfill\eject}
     
\def\endpaper                  
  {\endmode\vfill\supereject}

\def\endit
  {\endpaper\end}

     
\def\ref#1{Ref.[#1]}                   
\def\Ref#1{Ref.[#1]}                   

\def\Equation#1{Equation [#1]}         
\def\Equations#1{Equations [#1]}       
\def\Eq#1{Eq. (#1)}                     
\def\eq#1{Eq. (#1)}                     
\def\Eqs#1{Eqs. (#1)}                   
\def\eqs#1{Eqs. (#1)}                   
\def\frac#1#2{{\textstyle{{\strut #1} \over{\strut #2}}}}

\def\sla{\raise.15ex\hbox{$/$}\kern-.57em}
\def\leaderfill{\leaders\hbox to 1em{\hss.\hss}\hfill}
\def\twiddle{\lower.9ex\rlap{$\kern-.1em\scriptstyle\sim$}}
\def\bigtwiddle{\lower1.ex\rlap{$\sim$}}
\def\gtwid{\mathrel{\raise.3ex\hbox{$>$\kern-.75em\lower1ex\hbox{$\sim$}}}}
\def\ltwid{\mathrel{\raise.3ex\hbox{$<$\kern-.75em\lower1ex\hbox{$\sim$}}}}
\def\square{\kern1pt\vbox{\hrule height 1.2pt\hbox{\vrule width 1.2pt\hskip 3pt
   \vbox{\vskip 6pt}\hskip 3pt\vrule width 0.6pt}\hrule height 0.6pt}\kern1pt}

%

%

%

%
\physgreek
%

\def\dsl{\raise.15ex\hbox{$/$}\kern-.57em\hbox{$\partial$}}
\def\nsl{\raise.15ex\hbox{$/$}\kern-.57em\hbox{$\nabla$}}
\def\gtwid{\,{\raise.3ex\hbox{$>$\kern-.75em\lower1ex\hbox{$\sim$}}}\,}
\def\ltwid{\,{\raise.3ex\hbox{$<$\kern-.75em\lower1ex\hbox{$\sim$}}}\,}
\def\undr{\raise.3ex\hbox{$\sim$\kern-.75em\lower1ex\hbox{$|\vec x|\to\infty$}}}

\def\[{\left [}
\def\]{\right ]}
\def\({\left (}
\def\){\right )}







\def\and{a^{\phantom\dagger}}

%
\def\id{\raise.72ex\hbox{$-$}\kern-.85em\hbox{$d$}\,}

\catcode`@=11
\newcount\r@fcount \r@fcount=0
\newcount\r@fcurr
\immediate\newwrite\reffile
\newif\ifr@ffile\r@ffilefalse
\def\w@rnwrite#1{\ifr@ffile\immediate\write\reffile{#1}\fi\message{#1}}

\def\writer@f#1>>{}
\def\referencefile{
  \r@ffiletrue\immediate\openout\reffile=\jobname.ref%
  \def\writer@f##1>>{\ifr@ffile\immediate\write\reffile%
    {\noexpand\refis{##1} = \csname r@fnum##1\endcsname = %
     \expandafter\expandafter\expandafter\strip@t\expandafter%
     \meaning\csname r@ftext\csname r@fnum##1\endcsname\endcsname}\fi}%
  \def\strip@t##1>>{}}

\def\citeall#1{\xdef#1##1{#1{\noexpand\cite{##1}}}}
\def\cite#1{\each@rg\citer@nge{#1}}	

\def\each@rg#1#2{{\let\thecsname=#1\expandafter\first@rg#2,\end,}}
\def\first@rg#1,{\thecsname{#1}\apply@rg}	
\def\apply@rg#1,{\ifx\end#1\let\next=\relax
\else,\thecsname{#1}\let\next=\apply@rg\fi\next}

\def\citer@nge#1{\citedor@nge#1-\end-}	
\def\citer@ngeat#1\end-{#1}
\def\citedor@nge#1-#2-{\ifx\end#2\r@featspace#1 
  \else\citel@@p{#1}{#2}\citer@ngeat\fi}	
\def\citel@@p#1#2{\ifnum#1>#2{\errmessage{Reference range #1-#2\space is bad.}%
    \errhelp{If you cite a series of references by the notation M-N, then M and
    N must be integers, and N must be greater than or equal to M.}}\else%
 {\count0=#1\count1=#2\advance\count1 by1\relax\expandafter\r@fcite\the\count0,%
  \loop\advance\count0 by1\relax
    \ifnum\count0<\count1,\expandafter\r@fcite\the\count0,%
  \repeat}\fi}

\def\r@featspace#1#2 {\r@fcite#1#2,}	
\def\r@fcite#1,{\ifuncit@d{#1}
    \newr@f{#1}%
    \expandafter\gdef\csname r@ftext\number\r@fcount\endcsname%
                     {\message{Reference #1 to be supplied.}%
                      \writer@f#1>>#1 to be supplied.\par}%
 \fi%
 \csname r@fnum#1\endcsname}
\def\ifuncit@d#1{\expandafter\ifx\csname r@fnum#1\endcsname\relax}%
\def\newr@f#1{\global\advance\r@fcount by1%
    \expandafter\xdef\csname r@fnum#1\endcsname{\number\r@fcount}}

\let\r@fis=\refis			
\def\refis#1#2#3\par{\ifuncit@d{#1}
   \newr@f{#1}%
   \w@rnwrite{Reference #1=\number\r@fcount\space is not cited up to now.}\fi%
  \expandafter\gdef\csname r@ftext\csname r@fnum#1\endcsname\endcsname%
  {\writer@f#1>>#2#3\par}}

\def\ignoreuncited{
   \def\refis##1##2##3\par{\ifuncit@d{##1}%
     \else\expandafter\gdef\csname r@ftext\csname r@fnum##1\endcsname\endcsname%
     {\writer@f##1>>##2##3\par}\fi}}

\def\r@ferr{\endreferences\errmessage{I was expecting to see
\noexpand\endreferences before now;  I have inserted it here.}}
\let\r@ferences=\references
\def\references{\r@ferences\def\endmode{\r@ferr\par\endgroup}}

\let\endr@ferences=\endreferences
\def\endreferences{\r@fcurr=0
  {\loop\ifnum\r@fcurr<\r@fcount
    \advance\r@fcurr by 1\relax\expandafter\r@fis\expandafter{\number\r@fcurr}%
    \csname r@ftext\number\r@fcurr\endcsname%
  \repeat}\gdef\r@ferr{}\endr@ferences}


\let\r@fend=\endpaper\gdef\endpaper{\ifr@ffile
\immediate\write16{Cross References written on []\jobname.REF.}\fi\r@fend}

\catcode`@=12

\citeall\refto		
\citeall\ref		%
\citeall\Ref		%

\catcode`@=11
\newcount\tagnumber\tagnumber=0
     
\immediate\newwrite\eqnfile
\newif\if@qnfile\@qnfilefalse
\def\write@qn#1{}
\def\writenew@qn#1{}
\def\w@rnwrite#1{\write@qn{#1}\message{#1}}
\def\@rrwrite#1{\write@qn{#1}\errmessage{#1}}
     
\def\taghead#1{\gdef\t@ghead{#1}\global\tagnumber=0}
\def\t@ghead{}
     
\expandafter\def\csname @qnnum-3\endcsname
  {{\t@ghead\advance\tagnumber by -3\relax\number\tagnumber}}
\expandafter\def\csname @qnnum-2\endcsname
  {{\t@ghead\advance\tagnumber by -2\relax\number\tagnumber}}
\expandafter\def\csname @qnnum-1\endcsname
  {{\t@ghead\advance\tagnumber by -1\relax\number\tagnumber}}
\expandafter\def\csname @qnnum0\endcsname
  {\t@ghead\number\tagnumber}
\expandafter\def\csname @qnnum+1\endcsname
  {{\t@ghead\advance\tagnumber by 1\relax\number\tagnumber}}
\expandafter\def\csname @qnnum+2\endcsname
  {{\t@ghead\advance\tagnumber by 2\relax\number\tagnumber}}
\expandafter\def\csname @qnnum+3\endcsname
  {{\t@ghead\advance\tagnumber by 3\relax\number\tagnumber}}
     
\def\equationfile{%
  \@qnfiletrue\immediate\openout\eqnfile=\jobname.eqn%
  \def\write@qn##1{\if@qnfile\immediate\write\eqnfile{##1}\fi}
  \def\writenew@qn##1{\if@qnfile\immediate\write\eqnfile
    {\noexpand\tag{##1} = (\t@ghead\number\tagnumber)}\fi}
}
     
\def\callall#1{\xdef#1##1{#1{\noexpand\call{##1}}}}
\def\call#1{\each@rg\callr@nge{#1}}
     
\def\each@rg#1#2{{\let\thecsname=#1\expandafter\first@rg#2,\end,}}
\def\first@rg#1,{\thecsname{#1}\apply@rg}
\def\apply@rg#1,{\ifx\end#1\let\next=\relax%
\else,\thecsname{#1}\let\next=\apply@rg\fi\next}
     
\def\callr@nge#1{\calldor@nge#1-\end-}
\def\callr@ngeat#1\end-{#1}
\def\calldor@nge#1-#2-{\ifx\end#2\@qneatspace#1 %
  \else\calll@@p{#1}{#2}\callr@ngeat\fi}
\def\calll@@p#1#2{\ifnum#1>#2{\@rrwrite{Equation range #1-#2\space is bad.}
\errhelp{If you call a series of equations by the notation M-N, then M and
N must be integers, and N must be greater than or equal to M.}}\else%
 {\count0=#1\count1=#2\advance\count1
 by1\relax\expandafter\@qncall\the\count0,%
  \loop\advance\count0 by1\relax%
    \ifnum\count0<\count1,\expandafter\@qncall\the\count0,%
  \repeat}\fi}
     
\def\@qneatspace#1#2 {\@qncall#1#2,}
\def\@qncall#1,{\ifunc@lled{#1}{\def\next{#1}\ifx\next\empty\else
  \w@rnwrite{Equation number \noexpand\(>>#1<<) has not been defined yet.}
  >>#1<<\fi}\else\csname @qnnum#1\endcsname\fi}
     
\let\eqnono=\eqno
\def\eqno(#1){\tag#1}
\def\tag#1$${\eqnono(\displayt@g#1 )$$}
     
\def\aligntag#1\endaligntag
  $${\gdef\tag##1\\{&(##1 )\cr}\eqalignno{#1\\}$$
  \gdef\tag##1$${\eqnono(\displayt@g##1 )$$}}

\def\eqalignno#1{\displ@y \tabskip\centering
  \halign to\displaywidth{\hfil$\displaystyle{##}$\tabskip\z@skip
    &$\displaystyle{{}##}$\hfil\tabskip\centering
    &\llap{$\displayt@gpar##$}\tabskip\z@skip\crcr
    #1\crcr}}
     
\def\displayt@gpar(#1){(\displayt@g#1 )}
     
\def\displayt@g#1 {\rm\ifunc@lled{#1}\global\advance\tagnumber by1
        {\def\next{#1}\ifx\next\empty\else\expandafter
        \xdef\csname @qnnum#1\endcsname{\t@ghead\number\tagnumber}\fi}%
  \writenew@qn{#1}\t@ghead\number\tagnumber\else
        {\edef\next{\t@ghead\number\tagnumber}%
        \expandafter\ifx\csname @qnnum#1\endcsname\next\else
        \w@rnwrite{Equation \noexpand\tag{#1} is a duplicate number.}\fi}%
  \csname @qnnum#1\endcsname\fi}
     
\def\ifunc@lled#1{\expandafter\ifx\csname @qnnum#1\endcsname\relax}
     
\let\@qnend=\end\gdef\end{\if@qnfile
\immediate\write16{Equation numbers written on []\jobname.EQN.}\fi\@qnend}
     
\catcode`@=12
\callall\Equation
\callall\Equations
\callall\Eq
\callall\eq
\callall\Eqs
\callall\eqs


\referencefile

\twelvepoint\doublespace

\title{An Orbitally Degenerate Spin Fluctuation Model for Heavy Fermion
Superconductivity}

\author{M. R. Norman}
\affil
Materials Science Division
Argonne National Laboratory
Argonne, IL  60439

\abstract

In this paper, a generalization of standard spin fluctuation theory is
considered by replacing the simple Hubbard interaction by the screened
Hartree-Fock
interaction for f electrons.  This model is then used in both an
LS and a JJ coupling scheme to construct the particle-particle scattering
vertex in an on-site approximation.  This vertex is shown to lead to an
instability for a superconducting pair state which obeys Hunds rules,
with L=5, S=1, and J=4.  The degeneracy of this state is broken by
anisotropy of the quasiparticle wavefunctions.  Detailed calculations
are presented for the case of $UPt_3$.

\bigskip

\noindent PACS numbers:  74.20.-z, 74.70.Tx

\endtitlepage

After over a decade's worth of theoretical work, there is no overall agreement
on a microscopic theory for heavy fermion superconductivity.  The overall
prejudice, though, is that the underlying pairing mechanism is similar to that
operative in superfluid $^3He$.  Most attempts at a theory based on this
approach have been to make the simplest possible modifications to the
standard single
orbital Hubbard interaction used in the $^3He$ problem.  These attempts have
had mixed success.  The philosophy of this paper will be to actually do for
heavy fermions what was done for $^3He$, that is to use the Hartree-Fock
interaction between f electrons including full orbital and
spin-orbital effects and construct the effective particle-particle vertex
by the appropriate diagrammatic summation.  In principle, this theory
contains all relevant physics within a spin fluctuation based approach.
Even at the simplest level, new physics emerges which is not present when
using a simple Hubbard interaction.  In particular, the maximum instability
in the particle-particle vertex occurs for a pair state which obeys Hunds
rules.  For f electrons, this corresponds to a state which has L=5, S=1, and
J=4.  The degeneracy of this multiplet is broken in real metals by
crystalline anisotropy effects in the normal state.  This is reflected by
(1) the orbital and momentum dependence of the bare susceptibility bubble
which forms the internal lines of the vertex and (2) the orbital and momentum
dependence of the quasiparticle wavefunctions which form the external lines.
In this paper, (1) is treated in a simple manner and (2) is treated within
a band theoretic approximation.  The frequency dependence of (1) and (2),
which acts to set the overall scale for $T_c$, is also treated in a simple
fashion.  The theory has the advantage that it can be systematically improved
by removing these approximations.  The above ideas are illustrated by
calculations for $UPt_3$.

In the first section, a motivation of this theory is given by looking at some
systematics of heavy fermion superconductors and by comparing the heavy fermion
problem to that of $^3He$.  In the second section, the general formalism is
described.
The single orbital version of this theory is shown to yield the paramagnon
model for $^3He$.  The standard spin fluctuation models worked on previously
are then shown to be a lattice generalization of the single orbital model.
In the third section, the formalism for the f electron problem is derived,
with the particle-particle vertex equations solved for various
approximations for the susceptibility bubble in both an LS and a JJ coupling
scheme.  General properties of the vertex are described based on group theory.
In the fourth section, the pair vertex is projected onto the Fermi surface.
Calculations are then described for the case of $UPt_3$ utilizing information
from a relativistic band structure calculation.  In the last section,
future directions, including the question of inter-site pairing effects,
will be discussed.  A shorter version of this work has appeared
earlier.\refto{nprl}

\bigskip

\noindent I.  Introduction

Sufficient evidence has accumulated over the past eleven years
to demonstrate that the superconductivity seen in a number of f electron
metals with large effective mass is unconventional in nature, that is, the
group representation describing the order parameter is almost certainly
not the identity representation ($\Gamma_1^+$).  This, along with a
variety of other facts, casts doubt on a traditional electron-phonon
mechanism as mediating the pairing.  The first theoretical work in this
area ten years ago showed a close connection of these metals with
superfluid $^3He$.\refto{and1}  In particular, they are near both a
magnetic and a
localization instability.  A classic example is $UPt_3$.  Doping with $Pd$,
for instance, causes this metal to become strongly antiferromagnetic.  Further
doping causes the f electrons to become localized.\refto{visser}
Anderson\refto{and1}
also emphasized that the on-site part of the interaction must be
playing a major role given the large ratio ($\sim$0.1) of the
superconducting transition
temperature to the Fermi energy.  This important observation
has been largely forgotten.  Anderson\refto{and2} was also the
first to point out that heavy fermion superconductors have two f atoms per
unit cell.  For on-site pairing, one can have an odd parity ground state in
this case (with one atom per cell, the pair state would have to be even
for on-site pairing).
This unusual observation has also been largely forgotten, except in a
later paper by Appel and Hertel\refto{ah} where a formalism for
describing localized pairs for $UPt_3$ was developed in great detail.
The reason the above points were largely forgotten was the observation
of antiferromagnetic spin fluctuations in several heavy fermion
superconductors by neutron scattering.\refto{aeppli}  Such
fluctuations occur because
of exchange interactions between near neighbor sites.  This led to a
picture of near neighbor pairing based on these fluctuations by a number
of authors.\refto{miyake,norm1}  Subsequent work largely concentrated
on generalizing these simple theories to handle the non-symmorphic
(HCP) lattice structure of $UPt_3$.\refto{pj,norm2}  These theories have
had mixed success.  In particular, available data on $UPt_3$ point to
a pair state from a two dimensional group representation with both
line and point nodes and probably of odd parity.\refto{norm,js}  This state
would have $\Gamma_6^-$ ($E_{2u}$) symmetry.  Although non-trivial group
representations occur in these calculations, this particular
state has never emerged as the ground state.  The last work done in
this area by the author\refto{norm3} indicated that the anisotropy of the
quasiparticle wavefunctions plays a fundamental role
because this problem cannot be reduced to an effective one-band form
given the two f atoms per unit cell.  Therefore, simple theories as pursued
above will simply
be inadequate for describing real heavy fermion metals and any results
generated by them questionable.

This paper is an attempt to break this theoretical deadlock.  This work
was motivated by several additional issues than those listed above.
A number of alternate theories have been proposed recently,
in particular by Cox\refto{cox1} and by Coleman et al,\refto{cole} which
emphasize an on-site pairing viewpoint.  Cox's work is important in
that he emphasized the important role that orbital effects play in
this problem.  Another key motivation was an experiment by Osborn
et al\refto{osb} which detected excitations between Coulomb multiplets
with high energy neutron scattering, not only in localized f metals like
$Pr$ and $UPd_3$, but also in $UPt_3$ itself.  This indicates that multiplet
correlations present in atoms survives even in a metal with itinerant f
quasiparticles.  In Table 1, a list of the seven known heavy fermion
superconductors are listed.  There are two striking things about this
table.  First, six of the seven known heavy fermion superconductors
are uranium alloys.  Moreover, there is strong experimental evidence
that the uranium atoms are close to an $f^2$ configuration.  The
magnetic susceptibilities of $UPt_3$\refto{visser} and
$UPd_2Al_3$\refto{steglich}
look almost identical to that of the local $f^2$ metal
$PrNi_5$.\refto{prni5}  The susceptibility of $URu_2Si_2$ has been
most successfully explained based on an $f^2$ ground state.\refto{urs}
Cox's quadrupolar model for $UBe_{13}$\refto{cox2} is also based on
an $f^2$ configuration.  It should also be noted that $UPt_3$ is very
similar to $UPd_3$ (similar crystal structures, almost identical f atom
separations) yet the latter is clearly a local $f^2$ metal.\refto{buyers}
In the high energy neutron data,\refto{osb} the Coulomb excitation
seen in these two metals looks very similar.  This would be hard to
accept if $UPt_3$ was not close to being $f^2$.  As for $CeCu_2Si_2$,
it may not be like the rest, although it has been pointed out that an
$f^2$ admixture is needed to explain its properties with the Anderson
impurity model.\refto{coxpc}  The importance of these facts is that since
the f atom has two bare f electrons per site, this leads to a strong
motivation that the superconducting pairs have two f quasiparticles
per site from a trial wavefunction point of view.  The second striking
point of Table 1
is that all of these metals either have two f atoms per unit cell,
or undergo some sort of magnetic transition at temperatures above the
superconducting
transition which gives a new unit cell with two f atoms.  As discussed
above, this fact has little relevance to a near neighbor pairing model
(since an atom in any crystal structure has near neighbors),
but plays a crucial role for on-site pairing (since one can have even
parity or odd parity pairing depending on the phase of the order parameter on
the two sites).

A further motivation of the importance of on-site pairing can be obtained
by comparing the case of $^3He$ to uranium alloys.  In Figure 1, a plot
is shown of the interaction potential of two He atoms.\refto{aziz}
This is amazingly
similar to what one would expect of two f electrons on a uranium
site.  In particular, there is strong repulsion at small interparticle
separation due to the Coulomb repulsion between the two f electrons,
there is attraction at intermediate distances (or order 3 a.u.) since the
ion core attraction exceeds this repulsion in this range (which leads to
an $f^2$ ground state), and then the potential weakens at large
separation due to the exponential decay of the f electron radial function.
The direct interaction potential for $^3He$, though, has been shown to
be inadequate for describing the pairing of He atoms in the superfluid
state (it predicts L=2 pairing).\refto{old}
The reason for this is the major role that collective effects play due
to polarization of the medium.\refto{bs,layzer,ab,naka,leggett}
This led to the development of a paramagnon
model for $^3He$.  In this model, a much simpler direct
interaction potential is used, a repulsive contact interaction
between atoms of opposite spin.  But this potential in turn is used to
sum a diagrammatic series to all orders, thus including the important collective
effects.  This gives a good description of the superfluid state of
$^3He$.\refto{ab,lvr}  In the next section, a generalization of this model is
considered which includes orbital interaction effects necessary in
dealing with f electrons.  It is then shown that the paramagnon model
for $^3He$, as well as previous spin fluctuation models for heavy
fermions, are subsets of this more general theory.

\bigskip

\noindent II.  General Formalism

The particle-particle vertex is defined by
$$
\Gamma^{abcd} = \Gamma_0^{abcd} - \sum_{e,f}
\Gamma_0^{aecf}\chi_0^{ef}\Gamma^{fbed}
\eqno(1)
$$
where $\Gamma_0$ is the bare vertex, $\chi_0$ is the bare susceptibility
bubble, the indices label orbitals, and the minus sign is due to the closed
Fermion loop defining the bubble (in $\Gamma$, the first two indices label
incoming lines, the last two outgoing lines).  $\Gamma_0$ is taken to be
the antisymmetrized Coulomb interaction ($V^{abcd}-V^{abdc}$)
$$
\Gamma_0^{abcd} = \sum_k c_k^{abcd} F_k
\eqno(2)
$$
where $c_k$ are combinations of 3j symbols and $F_k$ are Coulomb multipole
(Slater) integrals defined on page 217 of Condon and Odabasi.\refto{co} (A
simpler expression of this type has been used in earlier spin fluctuation
work.\refto{doniach})
For s electrons, Eq. 2 reduces to
$$
\Gamma_0^{abcd} = \delta_{ac}\delta_{bd}-\delta_{ad}\delta_{bc}
\equiv {1 \over 2}
(\delta_{ac}\delta_{bd}-\vec\sigma_{ac} \cdot \vec\sigma_{bd})
\eqno(3)
$$
where the last term is a scalar product of Pauli spin matrices and the
indices now label just spins (1 for up, 2 for down).  Eq. 1 is easily
solved, giving\refto{ab,naka}
$$
\Gamma^{1111} = -F_0^2\chi_0/(1-F_0^2\chi_0^2)
\eqno(4)
$$
$$
\Gamma^{1212} = F_0 + F_0^3\chi_0^2/(1-F_0^2\chi_0^2)
\eqno(5)
$$
Eq. 4 (5) is a sum of odd (even) number of longitudinal (e=f) bubbles.  The
triplet (S=1) vertex is just Eq. 4, the singlet (S=0) vertex is $2\Gamma^{1212}
-\Gamma^{1111}$ (since $\Gamma^{1212}$ is half the sum of the singlet and
triplet vertex\refto{ab}).  Alternately, the singlet vertex is
$\Gamma^{1212}-\Gamma^{1221}$ (i.e., antisymmetrizing Eq. 5) where the latter
term is a sum of transverse (e$\neq$f) bubbles (i.e., ladder diagrams).  These
expressions are equivalent of course.  Eqs. 4 and 5 form the basis of the
standard paramagnon model (where $F_0$ is generally denoted as I).

The first thing to note from the above is that the triplet vertex is negative
(attractive) and the singlet one positive (repulsive).  Thus, the vertex
exhibits Hunds first rule (maximal S).  In the $^3He$ problem, the L of the
pair state is determined
by projecting Eq. 4 onto the Fermi surface.  Since it is isotropic, and L
must be odd (since S=1), it is necessary to include the momentum dependence
of $\chi_0$ to obtain a non-zero projection.  As expected, L=1 is found since
this function has the largest projection on a spherical Fermi surface for
odd L harmonics.  Note that the momentum dependence is not critical, it is
only neccessary to give a non-zero projection.  In fact, very different
models of the momentum dependence of the vertex give identical pairing
coupling constants.\refto{lv2}  A physical picture of the pairing in these
models based on mutual interaction of the two particles via their polarization
clouds has been given by Leggett.\refto{leggett}

This picture can be contrasted with that given for the heavy fermions and
high $T_c$ cuprates based on singlet (L=2) pairing.\refto{miyake}
This "violation" of
Hunds rules is obtained by considering pairing of electrons on near neighbor
atoms in a nearly antiferromagnetic metal.  In such a case, $\chi(\vec Q)
 > \chi(0)$ where $\vec Q$ is the ordering wavevector (assumed to be
commensurate
with the lattice corresponding in real space to anti-alignment of near
neighbor spins) and $\chi$ is the dressed bubble.  This means in real
space that $\chi(R,R')$ is negative ($R,R'$ are site indices, with $R'$ a near
neighbor of $R$).
This can be achieved by having momentum dependence in either $\chi_0$
or in $\Gamma_0$ ($\chi$ being defined by an equation similar to Eq. 1).
The latter is preferable, in that commensurate Q are rarely obtained for
$\chi_0$ except in special circumstances, and has been used for fitting neutron
scattering data both in heavy fermions\refto{aeppli} and in high $T_c$
cuprates.\refto{ruln}  Now, the same thing which gives a negative
$\chi(R,R')$ also gives a sign for $\Gamma(R,R',R,R')$ opposite to that
of $\Gamma(R,R,R,R)$.  Therefore, if the order parameter is such
that $\Delta(R,R)$ is zero and $\Delta(R,R')$ is non-zero (such as found for
certain d-wave states), then one can have S=0 pairing.  Note the complete
difference in physics than that discussed above for the $^3He$ problem.
In particular, the momentum dependence is crucial for this argument.
From the lattice point of view, $^3He$ is actually more closely related to
on-site pairing models than to near neighbor pairing models.

\bigskip

\noindent III.  f Electrons

The formalism of the previous section is now applied to the problem of f
electrons.  At the bare interaction level, we are already faced with the
problem that the uranium ion is in the intermediate coupling regime (i.e.,
midway between LS and JJ coupling).\refto{gold}  On the other hand,
the spin-orbit interaction is large enough that in electronic
structure calculations, no j=7/2 quasiparticles are occupied.\refto{abc}
Because of this, the susceptibility bubble $\chi_0$ will be almost pure
j=5/2 in character.  Therefore, even if the bare interaction is in the
intermediate coupling regime, the effective interaction for quasiparticles
which comes out
of Eq. 1 will be in the JJ coupling limit.  Despite this, we will
start out by deriving results in the LS coupling limit to make connections
to the $^3He$ problem discussed in the last section.  Then, we will turn
to the JJ scheme.

We start by reviewing the multiplet structure of the $f^2$ uranium ion,
shown in Figure 2.\refto{gold}  There are 3 spin triplets (L=1,3,5, each
spin-orbit split into 3 orbital multiplets) and 4 spin singlets
(L=0,2,4,6).  This level
structure can be fit by the following scheme.  At the Hartree-Fock level,
only Slater integrals of even rank appear.  Fits of the spectra can be
achieved by reducing these integrals on average by 38\%.  This effect is
due to screening caused by Coulomb correlations (i.e., the particles try
to avoid one another, thus reducing their effective interaction) which can be
approximately
calculated within a configurational interaction (CI) scheme.  In addition,
CI causes effective operators of odd rank to appear not present at the
Hartree-Fock level (known as Trees parameters).  These terms are rather
small, though, and we ignore them.  To discuss the level scheme, it is
useful to find linear combinations of the $c_k$ coefficients of Eq. 2
which more clearly reflect the group theoretical structure of the f
electrons, which are labeled $e_k$.
This has been achieved by Racah\refto{racah,judd,co} and is
equivalent to replacing the Slater integrals $F_k$ (k=0,2,4,6) by
linear combinations $E_k$ (k=0,1,2,3).
$E_0$ is defined such that all $f^2$ terms have this energy ($e_0$=1).
It is equal to $F_0$ plus a linear combination of the
other $F_k$ terms and is equivalent to the Hubbard U parameter (the
other $E_k$ parameters do not contain $F_0$).  $E_1$
is defined so as to distinguish spin singlets from spin triplets, with
$e_1$=0 for triplets and $e_1$=2 for all singlets but L=0.  Note that these
coefficients are identical to the s-electron problem of the previous
section.  Thus, $E_1$ plays the same role in the f electron problem as the
paramagnon I ($F_0$) plays in the $^3He$ problem.  This is quite interesting,
since $E_1$ is a shape fluctuation term instead of a charge fluctuation
term (i.e., it does not involve $F_0$).
This means that the inclusion of orbital degeneracy qualitatively changes
the physics relative to single orbital models.  As for the L=0 singlet,
it has $e_1$=9.  This is another consequence of orbital degeneracy, basically
L=0 has a different group structure than L=2,4,6 since it already appears
at the $f^0$ level (i.e., it has a different "quasi-spin" or "seniority").
We summarize by writing down the expression for $e_1$
$$
e_1 = <LS|q_{12}+{1 \over 2} -2\vec s_1 \cdot \vec s_2|LS>
 = 2 - S(S+1) + 7\delta_{L0}
\eqno(6)
$$
where 1,2 label the two electrons, q is the seniority operator, and $\vec s$
is the spin operator.  Note the similarity to the last expression of Eq. 3.
As for $e_2$, it is isomorphic to L=2,4,6 and thus acts to split these
three multiplets apart (it is zero for all other L states).  The expression
for this term is quite complicated and will not be written down.
The most interesting parameter is $E_3$.  It acts
to split apart the three spin triplets, with $e_3$=-9,0,33 for L=5,3,1
respectively (it is non-zero for all terms but L=0).  This can be written as
$$
e_3 = [{1 \over 2}-S(S+1)][L(L+1)-24g(U)]
\eqno(7)
$$
where g(U) is the Casimir operator of the group $G_2$ with U labeling the
representation of $G_2$ appropriate for a particular LS state (note that
L(L+1) is the Casimir operator for the group $SO_3$).  The interest is that
this has similarities to the case of p electrons, where VanVleck\refto{vv}
showed
long ago that the interaction between two p electrons can be written as
$$
w_{12} = F_0 + (-5 - 3\vec l_1 \cdot \vec l_2 - 12\vec s_1 \cdot \vec s_2)F_2
\eqno(8)
$$
with $\vec l$ the orbital angular momentum operator.  This looks like an orbital
generalization of the last expression in Eq. 3.  A summary of the $e_k$
coefficients for the $f^2$ states is given in Table 2.

We now turn to a solution of Eq. 1.  At this time, we will assume that
$\chi_0^{ef} = \chi_0$ (i.e., no orbital dependence to the bubble).  This
approximation will be discussed below.  States of definite LS
have antisymmetrized wavefunctions.  So, the vertex for the state L=5,
$M_L$=5 (S=1, $M_S$=1) will be
$$
\Gamma_{5,5} = \Gamma^{3232} - \Gamma^{3223} - \Gamma_0^{3232}
\eqno(9)
$$
where indices label $m_l$ (with all spins up).
The first term involves longitudinal bubbles, the second transverse
bubbles, and
the last compensates for double counting (since the bare vertex is
antisymmetric by definition).  Vertices for other LS states can be
obtained by either using the appropriate antisymmetric combination of
the $\Gamma$ or by employing Slater's diagonal sum rule.\refto{co}
This is equivalent to the two forms of the singlet vertex
in the paramagnon model discussed after Eq. 5.

In certain cases, analytic results can be derived by expanding Eq. 1
into a bubble summation, just as was done in the previous section for
$^3He$.  In particular, let us start with just including
the $E_0$ term.  The result for all states of definite LS is
$$
\Gamma = E_0/[(1-E_0\chi_0)(1+13E_0\chi_0)]
 + E_0^2\chi_0/(1-E_0\chi_0)
\eqno(10)
$$
The first term comes from longitudinal bubbles, the
second from transverse bubbles and double counting.
Note that 13 is the orbital degeneracy (14) minus 1.  This expression, plotted
in Figure 3,
is always repulsive.  The behavior of this term is that as $\chi_0$ increases
from zero, the repulsion is reduced compared to $E_0$ then begins to increase
again and diverges at $E_0\chi_0$=1.  This corresponds to a localization
instability.

Now assume that only $E_1$ is non-zero.  Analytic results can also be obtained.
For the spin triplet states, Eq. 1 is now
$$
\Gamma = -11E_1^2\chi_0/[(1-81E_1^2\chi_0^2) (1-4E_1^2\chi_0^2)]
+ 2E_1^2\chi_0/(1-4E_1^2\chi_0^2)
\eqno(11)
$$
where the first term comes from longitudinal bubbles, the second from
transverse bubbles.  This expression, plotted in Figure 4, is zero for
$\chi_0$=0 and then has a
negative divergence as $9E_1\chi_0$ approaches 1, corresponding to a magnetic
instability.  This behavior is analogous
to the triplet vertex in $^3He$.  Eq. 1 can also be solved for the L=6
singlet
$$
\Gamma = (4E_1+13E_1^2\chi_0-126E_1^3\chi_0^2-162E_1^4\chi_0^3)
/[(1-81E_1^2\chi_0^2) (1-4E_1^2\chi_0^2)] - 2E_1
\eqno(12)
$$
where the first term comes from longitudinal bubbles, the second from
transverse bubbles, and the third from double counting.  Just as for the singlet
vertex in $^3He$, this interaction, plotted in Figure 4, is repulsive for
$\chi_0$=0 (equal to
2$E_1$) and has a positive divergence at $9E_1\chi_0$=1.  This expression
should also be valid for the L=2 and L=4 singlets.  The vertex for the
L=0 state would be more repulsive.

The general solution to Eq. 1 is difficult to construct analytically due to
the complicated nature of the bare vertex when all $E_k$ terms are included.
Solving numerically would also appear to be difficult since their are four
orbital indices involved.  Progress can be made, though, by defining
$$
\tilde\Gamma^{be} = \Gamma^{b+n,e,b,e+n}
\eqno(13)
$$
with this definition, Eq. 1 reduces to
$$
\tilde\Gamma^{be} = \tilde\Gamma_0^{be} - \sum_f \tilde\Gamma_0^{bf}
\chi_0^{f,f+n} \tilde\Gamma^{fe}
\eqno(14)
$$
Thus, an $N^4$ matrix equation has been reduced to an $N^2$ matrix equation
for each value
of n (n=0 are the longitudinal bubbles, the rest are transverse bubbles).
This is easily solved on the computer given input values for the $E_k$
and $\chi_0$.

The $E_k$ parameters were taken from Goldschmidt.\refto{gold}  They have
been fit to uranium ion data (the level scheme for $\chi_0$=0 in Figure 5
differs in some quantitative details from the experimental level scheme of
Figure 2 since
spin-orbit has not been included at this point).   These
parameters are on average 62\% of their Hartree-Fock values due to screening.
By comparing to the high energy neutron scattering data of Osborn et al on
$UPt_3$,\refto{osb} these parameters should be reduced by another 28\%
when going
into the solid.  The latter effect is only of quantitative significance and
is based on one transition which is seen (assumed to be from $^3H_4$
to $^3F_2$), so
is ignored in order to use the well established values of $E_k$ listed in
Goldschmidt's article.  Hopefully, detailed solid state values of these
parameters will become available with further experimental work.  An additional
note is that the fitted $E_0$ is referenced to some arbitrary value of the
energy zero, so has no intrinsic meaning.  The value listed in the Goldschmidt
article, though, gives an $F_0$ of 1.83 eV which is fortuitously close to
estimates of the screened Coulomb U for uranium,\refto{herbst} so we retain
it without adjustment.

A final comment concerns the energy zero of the problem.
Superconductivity involves an instability of the Fermi surface.  For a
uranium ion, two f electrons are occupied.  Even in band structure
calculations for $UPt_3$, the
number of occupied j=5/2 f electrons is just above two.  Thus, the
term $E_0$ (the Coulomb repulsion between the two f electrons) is already
included in the definition of Fermi energy and
represents the
zero of energy for the uranium problem (for the cerium case, where only one
f electron is occupied, $E_0$ is not included in the energy zero, since
it represents the energy of $f^2$ above $f^1$).
But, the energy term for the L=5 ground state of an $f^2$ ion is $E_0-9E_3$.
What about the term $E_3$?  Since
this term cannot be reduced to an effective single particle form, it would
not seem to enter into the definition of the quasiparticle Fermi energy.
This gives the rather bizarre result that the L=5 vertex is already attractive
at the bare interaction level.  Of course, one could imagine a scenario
where one considered an $f^2$-$f^3$ Anderson lattice model, with the effects
of $E_3$ built into the ground states.  The effective quasiparticle operators
in this case might implicitly contain the effects of $E_3$.  Since a detailed
theory of this has not been worked out yet, we cannot make any definitive
conclusions one way or the other.  Since the effective interaction, though,
strongly departs from the bare value as $\chi_0$ increases from zero,
this question is of minor significance.  For purposes of this paper, we
assume that the energy zero is at $E_0$ for the $f^2$ case and 0 for the
$f^1$ case.

In Figure 5, $\Gamma$ is plotted for the triplet states L=5, L=3, and L=1
and for the singlet L=6.  The other singlets will have similar behavior.
Just as found for the $^3He$ problem, the triplet interactions are attractive
and the singlet ones repulsive, with an instability in both cases at
$(E_0+9E_1)\chi_0$=1.  Note this criteria is a combination of the two analytic
results discussed above, thus the instability has both a localization and
a magnetic component.  This observation indicates that the debate concerning
both $^3He$ and heavy fermions about whether the physics is nearly localized
or nearly magnetic is merely semantics, as both effects are intertwined.
A significant difference from the $^3He$ case is the effect
of orbital interactions in the current problem.  In $^3He$, the orbital
degeneracy of the pair state is lifted by Fermi surface projection of the
vertex; in the
f electron case, this degeneracy is already lifted by the interaction itself.
Note that the largest attractive instability is for L=5, S=1 (in such a
state, the Coulomb repulsion is minimized).  Thus, the pair
state is predicted to satisfy both Hunds first and second rules, and is a
generalization of the results obtained for $^3He$.

We now turn to a discussion of the problem at the JJ coupling level, which
as argued above, is more physically relevant at the quasiparticle level than
the LS scheme (or even the intermediate coupling scheme).  The bare interaction
vertex can be gotten by replacing the orbital-spin indices of Eq. 2 by
the indices $\mu$ which range from -5/2 to 5/2 (we assume only j=5/2
quasiparticles are involved).  By taking into account the mixed spinor nature
of the relativistic orbitals, the
formulas for $c_k$ can be calculated in a manner similar to that used
on p. 217 of Condon and Odabasi.\refto{co}  Note that only k=2 and 4 are
involved (k=6 comes in when considering j=7/2 states).  The resulting
expressions were checked against tabulated results on page 560 of de-Shalit
and Talmi.\refto{shalit}  As in the LS case, the $F_k$
are not useful in exploiting the group properties of the f electrons, so one
rotates to another basis $E_k$.  It should be noted that these $E_k$ are not
the same $E_k$ as in the LS case.  They have been discussed in the context
of nuclear physics, where JJ coupling has been traditionally of more
use,\refto{flowers} and the analogous $e_k$ coefficients are listed in Table
3.  In this case, there are only 3 terms, J=4,2,0 (corresponding to L=5,3,1
in the LS case).  As in the LS case, $e_0$ is 1 for all states.  $e_1$ is
3 for J=0, 0 otherwise which has analogies to $e_1$ in the LS case.
In particular,
J=0 appears at the $f^0$ level and thus has a different quasi-spin (i.e.,
seniority) than J=2,4.  The term $e_2$ splits the latter two states apart
(similar to $e_3$ in the LS case).
In this notation, the ground state J=4 term is $E_0-5E_2$.  This is lower
than the $E_0$ energy zero just as found for the L=5 LS case.
By fitting these values for states of definite J, we can infer an interaction
vertex analogous to VanVleck's\refto{vv} of Eq. 8 for the p electron LS case
$$
w_{12} = E_0 + {q_{12} \over 2} E_1 + [-2 \vec j_1 \cdot \vec j_2 -
{5 \over 2} (1 + q_{12})] E_2
\eqno(15)
$$
where $q_{12}$ is a seniority operator ($<J|q_{12}|J>=6\delta_{J0}$) and
$\vec j$ is the total angular momentum operator.

Analogous analytic series can also be constructed.  In particular, keeping
just $E_0$ gives an expression like Eq. 10 with 13 replaced by 5 since the
orbital degeneracy is now 6 instead of 14.

We now solve Eq. 1 exactly as done for the LS case.  The interaction parameters
are again obtained from Goldschmidt.\refto{gold}  The results are
plotted in Figure 6.  An attractive instability is found for J=4, a repulsive
instability for J=2,0.  The instability occurs at a value
$(E_0+E_1+12E_2)\chi_0$=1.  This result is important in that it makes a very
definite prediction, if a paramagnon like picture analogous
to $^3He$ applies to heavy fermion superconductors, a pair state of relative
J=4 should be realized.

How does this change if we replace uranium by cerium?  First, the energy zero
does not contain $E_0$ so that the bare interaction is more repulsive.  Second,
$E_0$ is about 3 times larger since U is around 6 eV for cerium
ions.\refto{free}  Although
this means that the instability occurs for a smaller value of $\chi_0$, we
expect the interaction to be more repulsive since $E_0$ is larger.  This is
illustrated in Figure 7 which is analogous to Figure 6
except parameters tabulated
by Goldschmidt for the cerium ion are used\refto{gold2} (with an $E_0$ of
6.0 eV).  As can be seen, the interaction
in all channels is repulsive except very close to the instability for J=4.
On the other hand, in strong coupling calculations for the paramagnon model
in $^3He$, the calculated superfluid transition temperature actually turns
off as the instability is approached since the energy scale of the paramagnon
(proportional to $1-I\chi_0$) is going to zero (that is, the $T_c$ maximum
is near but not at the instability).\refto{lv}
Because of this, pairing is possible
for cerium alloys but much less likely than the uranium case where one finds
a larger range of $\chi_0$ where there is attraction.  An alternate
view is that a pair wavefunction with two f quasiparticles
has much
less overlap with the bare f ion wavefunction in the cerium case since the
$f^2$ component in cerium is much smaller than in uranium.  This is in accord
with experimental observations discussed in the first section.

We now discuss the issue of orbital and momentum dependence of the
susceptibility bubble.  In heavy fermion uranium alloys, there is not much
evidence for crystal field effects.  This indicates that all the j=5/2 f
orbitals are strongly mixed, as predicted by band theory (as discussed for
Kondo lattice models by Zwicknagl,\refto{zw}
if the Kondo temperature is larger than the crystal field
splittings, then all f orbital energies get renormalized to the Fermi energy
and are intermixed; this appears to be the case in $UPt_3$).  Because of this,
one would expect the orbital and momentum dependence of $\chi_0$
to be rather weak.  There are inter-site interactions, though, which
give the full susceptibility, $\chi$, momentum dependence.
An argument against the importance of this effect for the pairing has been
given by Anderson\refto{and1}
where he emphasizes the dominance of the on-site interaction
given the large ratio of $T_c$ to $E_F$.  The issue of intersite pairing
will be discussed
in the last section.  On the other hand, the susceptibility in metals like
$UPt_3$ is strongly dependent on field direction.  Whether this is
a consequence of crystal field effects or simply due to inter-site
correlations is an unresolved matter\refto{visser} although neutron
scattering data in $UPt_3$ point to the latter.\refto{aeppli}  If this
is a crystal field effect, it can be represented by $\chi_0$.  If $\chi$
is maximal for fields along the c axis (like in $URu_2Si_2$), this
indicates that the longitudinal bubbles are dominant.  If one redoes
Figure 6 with just longitudinal bubbles, then the J multiplets are split
into various $M_J$ terms (actually, those $M_J$ combinations which have
the appropriate crystal symmetry) with the maximum $M_J$ configuration
having the most attractive instability.  On the other hand, if $\chi$ is
maximal for fields in the basal plane (like in $UPt_3$) then
transverse bubbles with n=1 are dominant (n as defined in Eq. 13 and 14).
This in turn leads to the minimum $M_J$ configuration being preferred.  Higher
values of n would indicate the importance of quadrupolar (n=2) effects, etc.
These terms play an important role in certain theories.\refto{cox1,cox2}
These observations are summarized in Table 4 (where it should be noted
that a J=4 state is always preferred).  In this paper, these
effects will be further ignored although they are relatively easy to
incorporate if they can be accurately determined (and if they are indeed
due to $\chi_0$ itself).  An argument that these effects cannot be too
strong is that an $M_J$=0 pair state is not consistent with experimental
data in $UPt_3$ since it is a single dimensional group representation,
although similar anisotropic
spin fluctuation work in a spin-only approximation gave an analogous
$M_S$=0 pair state,\refto{norm2} which is consistent with observations
of anisotropy in the upper critical field.\refto{cs}  These issues will
be discussed further in the next section.  As will be discussed in the
next section, projection of the vertex on the Fermi surface will also lead
to lifting of the degeneracy of the J manifold (analogous to the lifting of
L degeneracy in $^3He$).  It is that effect we concentrate on in this paper.

We conclude this section with a discussion of the general vertex.
In $^3He$, the full vertex can be written in a form analogous to the
bare expression in Eq. 3, that is a density piece proportional to the
delta functions and a spin piece proportional to the scalar product of
spin operators.\refto{ab}  Given the similarity of Eq. 3 to that of, say,
Eq. 15, this should also be possible in the f electron case (as long as one
restricts to states of definite LS or J).  In particular,
there will be a density piece, a quasi-spin (seniority) piece, and an
orbital piece proportional to the scalar product
of total angular momentum operators.  This justifies some of the
phenomenological interactions used in previous work.\refto{norm3}  Further
exploitation of these ideas should give us a more fundamental insight into
the properties of the full vertex for f electrons.  We should note that
the screened Slater integrals $F_k$ can be considered as analogues of
the Landau F functions of Fermi liquid theory.

\bigskip

\noindent IV.  Application to $UPt_3$

We now wish to employ the formalism in the previous section to a real heavy
fermion superconductor.  We choose for this purpose $UPt_3$.  There are two
good reasons for this.  First, a variety of experimental data exist on
this metal which gives us a fairly good idea about what the order parameter is.
Second, extensive deHaas-vanAlphen data on
$UPt_3$\refto{lonz} give a Fermi surface in fairly good agreement with LDA band
structure calculations.\refto{nabc}  This indicates that the momentum dependence
of the LDA wavefunctions is fairly trustworthy.  The frequency dependence is
not, of course, since the effective mass in the measurements is about 16 times
the LDA band mass.  This is due to the fact that in heavy fermions, the self
energy has a large frequency derivative leading to a large mass enhancement.
Investigation of transport properties indicates the momentum derivative of the
self-energy must be rather weak so that the self-energy "rides" with the
Fermi energy.\refto{varma}  That is why the shape of the LDA Fermi
surface is about correct even though the mass is off by a large factor.
These issues are of importance since we want to convert the formalism of the
previous section to apply to quasiparticle states.  We can do this approximately
by taking the four bare external lines of the vertex in Eq. 1 and multiplying 
each of them
by the wavefunction renormalization factor $Z^{1/2}$, where as indicated above
1/Z$\sim$16.  This represents the effect that only Z of the bare electron is
in the quasiparticle pole.

To proceed further, note that
$$
<J,\alpha|\Gamma|J',\alpha'> = \delta_{JJ'}\delta_{\alpha\alpha'}
\Gamma_{J,\alpha}
\eqno(16)
$$
where $\alpha$ is a basis function of J which has the appropriate crystal
symmetry.  For axial symmetry, this would just be M.  For hexagonal symmetry,
they are listed in Table 5.\refto{ah}
The dependence of $\Gamma$ on $\alpha$ occurs
if anisotropy is put into $\chi_0$ as discussed in the last section.  Also,
for a multi-dimensional group representation, $\Gamma$ will be the same for
each $\alpha$ in the representation, unless the symmetry is lowered by some
external perturbation.  This has relevance for Ginzburg-Landau models of the
phase diagram for $UPt_3$ and will be discussed later.  For now, though, we
assume only a J dependence for $\Gamma$.  These are plotted in Figure 6.
Given this, we can now calculate the paring interaction on the Fermi surface.
We do this by constructing the product $|\vec k,-\vec k>$ and expanding this
in terms
of $|J,\alpha>$.  The first thing to note is that there are two degenerate
states for each $\vec k$ ($\vec k$, PT$\vec k$ where P is the parity operator
and T the time reversal one) and two for $-\vec k$
(P$\vec k$, T$\vec k$).\refto{and3}  The combination
${1 \over 2}(|\vec k,T\vec k>-|PT\vec k,P\vec k>)$ defines a pseudo-spin
singlet, $d_0$.  The three combinations ${1 \over 2}(|PT\vec k,T\vec k>
-|\vec k,P\vec k>)=d_x$,
$-{i \over 2}(|PT\vec k,T\vec k>+|\vec k,P\vec k>)=d_y$, and
${1 \over 2}(|\vec k,T\vec k>+|PT\vec k,P\vec k>)=d_z$ define a
pseudo-spin triplet, known as the d vector.
In the current theory, only the part of $|\vec k,-\vec k>$ on the same site is
involved in Eq. 16.  (Note that although the pair interaction is only attractive
for particles on the same site, the pairs are correlated out to a distance
of the coherence length, much like the problem of bound states of a potential
well where the particles spend most of their time outside the
well.\refto{leggett})  Now, the part of $|\vec k>$ involving j=5/2 states
is\refto{wf}
$$
|\vec k> = \sum_{\mu i} a_{\mu i}^{n\vec k} |\mu>_i
\eqno(17)
$$
where $\mu$ runs from -5/2 to 5/2, i is the f atom site index (1,2 for $UPt_3$),
and n is the band index (five bands are predicted to cross the Fermi energy in
$UPt_3$).  Thus, the coefficient of $|\vec k,-\vec k>$ involving
j=5/2 states
on the same site with the correct group representation structure for a
particular total J (J=0,2,4) is
$$
A_{\vec k}^{J \alpha j} = \hat P_{J \alpha j} \sum_{\mu \nu i}
a_{\mu i}^{n\vec k} a_{\nu i}^{n-\vec k}
\eqno(18)
$$
where j represents the pseudo-spin combination (0 for singlet, x,y,z for
triplet) and $\hat P$ is a projection operator which takes that part of the sum
which has the form of one of the basis functions (Table 5) with the appropriate
pseudo-spin combination discussed above.  Because of antisymmetry,
A changes sign from one site to
the other site in the unit cell for pseudo-spin triplets, and does not for
pseudo-spin singlets (for one f atom per cell, only pseudo-spin singlets
exist\refto{n1}).  Summarizing, the particle-particle vertex is
$$
<\vec k',-\vec k'|\Gamma|\vec k,-\vec k> = Z^2 \sum_{J \alpha j j'} \Gamma_J
A_{\vec k'}^{*J \alpha j'} A_{\vec k}^{J \alpha j}
\eqno(19)
$$
Since Eq. 19 is separable in $\vec k$ and $\vec k'$, this allows us to write
down the BCS coupling constant
$$
\lambda_{J \alpha} = N \Gamma_J Z^2 \sum_j
<|A_{\vec k}^{J \alpha j}|^2>_{\vec k}
\eqno(20)
$$
where N is the density of states, $< >_{\vec k}$ is an average over a
narrow energy
shell about the Fermi energy, and j runs over 0 for even parity, x,y,z for
odd parity.  For $UPt_3$, this average was done on a regular grid of 561
$\vec k$ points in the irreducible wedge (1/24) of the Brillouin zone,
keeping those n$\vec k$
states within 1 mRy of the Fermi energy (182 states total).  Those points
that are in symmetry planes of the zone are plotted in Figure 8.  The
number 1 mRy was chosen so as to have enough points to give a good
representation
of the Fermi surface with this size grid.  Note that in this model
$$
\Delta_{J \alpha j}(\vec k) \propto A_{\vec k}^{J \alpha j}
\eqno(21)
$$
where $\Delta$ is the order parameter.

In Table 6, coupling constants\refto{g5} for $UPt_3$ are shown modulo
$N\Gamma_JZ^2$ with the largest
occuring for J=0, $\Gamma_1^+$.
This is just the BW state with spin-orbit,\refto{bw} with a coupling constant
proportional to the square of the ratio
of the j=5/2 f density of states to the total density of states.  From Figure 6,
though, the $\lambda_{2\alpha}$ and $\lambda_{0\alpha}$ coupling constants
are repulsive and so do not play a role.
For the attractive J=4 case, the largest coupling constants are for odd parity
states.  This is because three pseudo-spin triplet terms contribute to Eq. 20
in this case as opposed to one pseudo-spin singlet for even parity states.
The importance of this is that pseudo-spin triplets only exist because of the
two f atoms per unit cell, which as illustrated in Table 1, all heavy fermion
superconductors have.  This is a property of an on-site pairing theory, and
has no relevance in near neighbor pairing models.  The largest coupling constant
occurs for $\Gamma_6^-$ ($E_{2u}$) symmetry, although several other states
have comparable sized coupling constants ($\Gamma_1^-,\Gamma_4^-$).
Note that this is an odd parity
two dimensional group representation.

Let us discuss this state.  For the odd parity case, the order parameter is
a vector.  At a general $\vec k$, all three components are involved because
of the
relativistic nature of the wavefunction coefficients.  For two dimensional
group representations, this means that the state will be non-unitary
($\vec d \times \vec d^* \ne 0$).  A discussion of the case of non-unitary d
vectors can be found in Sigrist and Ueda\refto{su} which play a major role in
a recent phenomenological theory of $UPt_3$.\refto{mach}  In the non-unitary
case, there are two gaps for each $\vec k$
$$
\Delta_{\sigma}(\vec k)^2 = |\vec d(\vec k)|^2+\sigma
|\vec d(\vec k) \times \vec d^*(\vec k)|^2
\eqno(22)
$$
where $\sigma$ is +/-.  Plots of this for the $\Gamma_6^-$ state are shown in
Figures 9 and 10 with the total density of states shown in Figure 11.  No
attempt has been made to fit these gaps to simple functions\refto{yip}
because the
complex momentum dependence of the wavefunctions would prohibit this.  Instead,
some general properties can be inferred.  For instance, the $d_z$ component
vanishes for $k_z$=0 as expected from earlier work.\refto{gt}  A surprise,
though, is that all three d vector components vanish on the zone face
$k_z$=$\pi$/c.  Thus, this state has a line node gap function, and provides
a counterargument to earlier statements that line node gap functions are not
possible for odd parity states.\refto{gt}  Moreover, all d vector components
vanish along the axis $k_x$=0,$k_y$=0 which gives rise to point nodes.  A
gap function of this sort (line nodes perpendicular to the c axis, point
nodes along the c axis) is consistent with a variety of experimental data
in $UPt_3$, including specific heat,\refto{sh} transverse ultrasound,\refto{tu}
penetration depth,\refto{pene} thermal conductivity,\refto{tc} NMR,\refto{nmr}
and tunneling\refto{pc} data.  Note that despite
the small value of the second gap in Figure 10, no "normal" component is
seen in the density of states of Figure 11.  Thus, this non-unitary state
differs from
the one considered by Machida et al\refto{mach} where one of the gap
components vanishes identically for all $\vec k$ so that there is a
normal component with half the value
of the normal state (in the theory of Coleman et al,\refto{cole} a similar
normal component occurs).  It should be remarked that although earlier specific
heat data\refto{sh} indicated a sizable normal component, this is probably
due to impurity effects since newer data do not show this
component.\refto{brison}
As for the two dimensional nature of the group representation,
our current understanding of the field-temperature-pressure phase diagram
of $UPt_3$ is in strong support of such a state.\refto{js,joynt,joynt3}
In particular,
weak magnetism is present which lowers the symmetry to orthorhombic.  This
acts to split the superconducting transition into two transitions.  Pressure
acts to eliminate both the magnetism and the splitting,\refto{hayden}
thus giving strong support for a two dimensional group representation, as
opposed to two nearly degenerate single dimensional group representations.
The main problem with this scenario is the presence of a term in
the gradient part of the free energy which tends to mix the two components of
the representation except for certain field directions.  This is in
contradiction to experiment which shows a fairly isotropic phase diagram with
respect to field direction (this was the main motivation for the nearly
degenerate model\refto{joynt2,garg}).
Sauls, though, has shown that for an axially symmetric Fermi surface
and axially symmetric basis functions, this mixing term is zero for
the $E_2$ representation.\refto{js}  In the general case, it is not, but its
mixing term is smaller than that of the $E_1$ representation.
This work will be discussed in another paper,\refto{vin} but it
suffices to say here that for the current theory, the mixing term for our
$\Gamma_6^-$ ($E_{2u}$) state is small enough so as to be promising in
regards to explaining the phase diagram.  Alternate theories based on two
nearly degenerate representations\refto{joynt2,garg,zl} are also consistent
with the current theory given the closeness of the coupling constants for
$\Gamma_6^-$, $\Gamma_1^-$, and $\Gamma_4^-$ (the latter state having the same
line node structure as $\Gamma_6^-$).

A final issue concerns the question of parity.  No change below $T_c$
for $UPt_3$ has been seen in the Knight shift\refto{knight} or induced moment
form factor,\refto{ff} indicating no change in the magnetic susceptibility
below $T_c$.  This is in support of an odd parity state, although one could
argue that in the heavy fermions, the quasiparticle (intraband) part of
the susceptibility is small compared to the VanVleck (interband) part, so
this conclusion is not definitive.  Choi and Sauls\refto{cs} have also shown
that the observed low temperature directional anisotropy of the upper critical
field\refto{hc2} is most easily explained with an odd parity pair state
with $M_S$=0
(note, the quantization axis is assumed to be along c).
Such a state came out of earlier non-relativistic spin fluctuation
calculations which took into account the directional anisotropy of the
susceptibility.\refto{norm2}  In the current case, though, spin and orbital
components are mixed and so a pure $M_S$=0 state is not possible.
On the other hand, the state found here, $|M_J|$=1, has
the largest projection of J on the basal plane of any two dimensional group
representation ($M_J$=0 has the largest projection, but is a singlet),
so is promising in that regard.  To
test this quantitatively would require a detailed calculation of the upper
critical field with both spin and orbital degrees of freedom taken into
account.  Certainly, if the current predicted state is correct, the $H_{c2}$
anisotropy cannot
just be a spin effect as envisioned by Choi and Sauls.  This can be seen
as follows.  A direct translation of their idea to the current state would
be to have a state of pure $d_z$ character in pseudo-spin space.  This can
be tested by rotating in pseudo-spin space at each $\vec k$ so that the
state $|\vec k>$ has maximal $J_z$ along the chosen quantization axis.
This was done for quantization axes along the a, b, and c axes of the
hexagonal crystal (for the a and b case, this lowers the system to
orthorhombic symmetry).  In all cases, the averages $<|d_i|^2>_{\vec k}$
were within 20\% to 30\% of each other,
i.e. there is no dominant $d_z$ component.  This is
consistent with the highly non-unitary nature of this state seen in Figures
9 and 10.  As indicated above, then, orbital effects in the upper critical
field must be playing an important role.  This would require a theory for
calculating $H_{c2}$ for a non-unitary multi-component d vector in the
strong spin-orbit coupling limit.

We finally turn to a discussion of $T_c$.  We should note that the density
of states already contains the renormalization factor 1/Z (i.e., N = $N_0$/Z
where $N_0$ is the bare density of states).  This means that
the prefactor $N\Gamma_JZ^2$ in Eq. 20 is analogous to the form
$\lambda_{\Delta}/(1+\lambda_Z)$ found in strong coupling theories\refto{lv}
since $N_0\Gamma_J$ would be the pairing coupling constant, $\lambda_{\Delta}$,
and 1/Z - 1 would be the mass coupling constant, $\lambda_Z$.  This is
consistent with the fact that renormalization of the external lines of the
vertex is a strong coupling effect.  On the other hand, the frequency
dependence of the bubble has not been kept, so one has to simulate this
by providing an energy cut-off of order the renormalized Fermi energy.
We note that the
size of the specific heat coefficient\refto{visser} and the neutron scattering
linewidth\refto{aeppli} are consistent with a renormalized energy scale
for $UPt_3$ of order 5 meV.  This is also consistent with estimating a band
structure Fermi energy and multiplying this by Z.  The last thing to be
determined is $\Gamma_J$.  This is made difficult by the fact that $\chi_0$
is being treated as a number in this paper, whereas in reality it is a higly
complicated function of momentum, frequency, and band and orbital indices.
Given the strong dependence of $\Gamma_J$ on
$\chi_0$ (and also the question of the energy zero), the exponential
dependence of $T_c$ on coupling constant, and
the uncertainties mentioned in the above approximations, the most illustrative
approach is just to see what value of $\chi_0$ is needed to obtain the
observed $T_c$.  With a $T_c$ of 0.5 K and a cut-off of 5 meV, this gives a
value for $\lambda$ of 0.205.  Since $\lambda=N\Gamma_4Z^2c$ where c from Table
6 for J=4, $\Gamma_6^-$ is 0.125, this gives a $\Gamma_4$ of -2.1 eV relative
to the energy zero.  From Figure 6, this gives a $\chi_0$ of 0.335, which is
0.92 of the divergence value, giving a Stoner renormalization of 12, comparable
to the mass renormalization value of 16 assumed from the beginning.  Since spin
fluctuation models based on the observed heavy fermion dynamic susceptibility
have a mass renormalization which goes like the Stoner factor, as opposed to
the log of the Stoner
factor which one gets for a Lindhard function,\refto{norm1} then there
is a overall consistency in these numbers.  This can be further demonstrated
by estimating $\chi$, obtained by multiplying $\chi_0$ by the Stoner
renormalization (12)
then by the square of the orbital degeneracy (36).  This gives a value of
about 150 states/eV, comparable to the 180 states/eV given by the specific
heat $\gamma$ of 420 mJ/mol $K^2$.\refto{visser}  Moreover, converting $\chi$
to proper units (with g=6/7 and j=5/2) gives 0.0075 emu/mol, comparable to data
from susceptibility measurements.\refto{visser}  To obtain
a more reliable estimate of these parameters would require doing a full strong
coupling calculation retaining the frequency dependence of the bubble.\refto{lv}

The large estimated size of $\Gamma_J$ of order 2 eV which must be renormalized
downwards by $Z^2$ might seem somewhat worrisome.  After all, wouldn't one
expect high $T_c$ in transition metals where Z is closer to one?  The question
of renormalizing the interaction downwards has been discussed by
Anderson\refto{and1} and reviewed by Lee et al.\refto{lee}  The main point
to emphasize
here is that the interaction parameters of this paper are only appropriate for
a system close to an $f^2$ configuration, and the same thing that is causing
the large value of $\Gamma_J$ is also causing a small value for Z.
In transition metals, the $E_k$
parameters are largely screened out and play no role.  Instead, one collapses
back to a standard spin fluctuation model with a Stoner interaction parameter
I.  Estimates based on this I give extremely low estimates of $T_c$, even in
palladium which has a large Stoner renormalization.\refto{pd}

\bigskip

\noindent V.  Future Directions

An advantage of the current approach is that the theory is systematically
improvable by removing various approximations made in this paper.  The most
severe of these is treating $\chi_0$ as a number.  A proper strong coupling
calculation would include the frequency dependence of this object.  This is
not too difficult if the simple relaxational form is used
$$
\chi_0(\omega) = {\chi_0\Gamma \over \Gamma-i\omega}
\eqno(23)
$$
where $\Gamma$ is the neutron scattering
linewidth.\refto{aeppli}  Work of this sort has been done in earlier spin
fluctuation models.\refto{norm1,norm2}  Of more
interest is the momentum and orbital dependence of this object.  The philosophy
of this paper is similar to that espoused early on by Anderson,\refto{and1}
that is the size of the effective interaction is large enough (large ratio of
$T_c$ to $E_F$) that the on-site interaction must play the central role.
This is also consistent with the observation that one of the defining properties
of the heavy fermion metals is their large f atom separation\refto{stewart}
giving weak inter-site interaction effects.  Even neutron scattering data
indicate that that part of the susceptibility where the bulk of the fluctuating
moment is has a relatively mild momentum dependence.\refto{aeppli}  On the
other hand, these effects played a crucial role in earlier spin fluctuation
models, so it is of interest to see how they would enter the current formalism.
In the real space approach taken here, these effects would be simulated by
adding a term $\chi_0(R,R')$ in addition to the term $\chi_0(R,R)$ where $R,R'$
are site indices with $R'$ a near neighbor of $R$.  Near neighbor effects
would show up in Eq. 1 at first order in this bubble.
At second order in this bubble, there would be terms which would
also affect on-site pairing ($\chi_0(R,R')\chi_0(R',R)$).  Assuming Eq. 1
can be solved, one is left with a vertex which contains on-site terms,
near neighbor terms, etc.  The general properties of this vertex along with
the functional forms of the near neighbor and next near neighbor pairs are
discussed in Appel and Hertel.\refto{ah}  One simply has to combine this work
with that one to get a complete solution.  It is complicated, but doable.
A simple argument, though, can be used against near neighbor pairing in
that the same sign difference between on-site and near neighbor interactions
in the antiferromagnetic case which gives rise to S=0 pairing in the single
orbital case (end of Section II) should give J=0 pairing
in the orbitally degenerate case, since from Figure 6, J=0 has
maximal repulsion for the on-site case.  Such a state is a singlet,
and is completely inconsistent with the data we have discussed on
$UPt_3$.

The other issue concerns orbital dependence of the bubble.  Calculations of
this sort exist in the
literature for $UPt_3$\refto{nof} and are rather tedious, as they involve
calculating matrix elements of relativistic wavefunctions over a fine enough
grid in k space to get reliable values of $\chi_0$.  As discussed in Section
III, one might get around this difficulty by simulating these effects with
some effective crystal field model.  The most probable picture based on the
temperature dependence of the
neutron scattering data, though, is that this effect enters most prominently in
the $\chi_0(R,R')$ term.\refto{aeppli}

The next issue concerns feedback effects which are important in the physics
of superfluid $^3He$.  In that case, all L=1 states are degenerate at $T_c$.
Therefore, one would expect the isotropic state, the BW state, to
have the lowest free energy.  On the other hand, the susceptibility changes
below $T_c$ which affects the pair interaction and favors states with
maximal anisotropy, giving rise to the ABM state.\refto{ab,leggett}
In the current
problem, though, (1) this degeneracy is already broken in the normal state
due to the crystal lattice and (2) there is no experimental evidence for a
change in the susceptibility below $T_c$ (based on Knight shift\refto{knight}
and induced moment form factor\refto{ff} measurements), although it should
be noted that the
interband, or VanVleck, component most likely dominates the
susceptibility.\refto{nof}  (Neutron scattering
experiments which access the low momentum, low frequency part of the dynamic
susceptibility would be helpful in extracting out the quasiparticle part of
the susceptibility and seeing how it changes below $T_c$.)
Because of this, feedback effects
probably do not play an important role in the heavy fermion problem (the
large ratio of $T_c$ to $E_F$ would argue against this, though).  A related
effect is whether the interaction changes as a function of field (this could
be connected to the $H_{c2}$ anisotropy discussed in the previous section).
Magnetization data look very linear in field for all field directions for
the field range of interest\refto{visser} which would argue against
this.  On the other hand, as $T_c$
depends exponentially on coupling constant, small changes in the quasiparticle
wavefunctions in an applied field could lead to noticable effects, especially
given the low effective Fermi energy.  This
could be simulated in the current theory by rediagonalzing the band structure
wavefunctions in the presence of the appropriate field and see how the
coupling constants listed in Table 6 change.

A related effect is the observed splitting of $T_c$ in $UPt_3$ that was
discussed in the previous section.  This has been treated in the past by
a phenomenological symmetry breaking field thought to be due to the
orthorhombic distortion associated with the weak antiferromagnetism.\refto{hess}
This could also be simulated in the current model by applying a weak
staggered field (the new orthorhombic cell would contain four uranium atoms)
and rediagonalizing the band structure wavefunctions, from which the splitting
of the coupling constants could be determined.

The author would like to conclude by saying that the physics of heavy fermion
superconductors is complicated enough (as this paper demonstrates) that the
picture offered here may not be complete.  On the other hand, he feels that
the ultimate theory for these materials must look in some form like what is
being proposed here, since the orbital dependence of the f electrons and
the short range nature of the interactions should play a crucial role.  It
is promosing that this model has certain qualitative features
reflected in the data (preference for uranium with two f atoms per unit cell)
which are hard to understand from earlier spin fluctuation theories.
Also, the predicted
pair state for $UPt_3$ has many promising aspects also missing in earlier
theories.  Moreover, there is a conceptual beauty to having a pair state
which has maximal L and maximal S, as this is a direct generalization of
the physics of $^3He$ with which heavy fermions share many qualitative
features.  Hopefully, with increased experimental and theoretical effort,
we can determine whether this is indeed the right approach to pursue for
solving this problem.

\bigskip 

This work was supported by U.S. Department of Energy, Office of Basic Energy
Sciences, under Contract No. W-31-109-ENG-38.  The author would
like to acknowledge the hospitality of the Physics Dept., Uppsala University,
where this work was begun, and to thank Seb Doniach, Kathyrn Levin,
and Jim Sauls for helpful discussions.

\vfill\eject

\references

\refis{lv} K. Levin and O.T. Valls, \prb 17, 191, 1978.

\refis{pd} D. Fay and J. Appel, \prb 16, 2325, 1977.

\refis{urs} G.J. Nieuwenhuys, \prb 35, 5260, 1987.

\refis{doniach} S. Doniach, \prl 18, 554, 1967; J.R. Schrieffer, \prl 19,
644, 1967.

\refis{norm3} M.R. Norman, \prb 48, 6315, 1993.

\refis{norm2} M.R. Norman, \prb 43, 6121, 1991; \prb 41, 170, 1990;
\prb 39, 7305, 1989; \prb 37, 4987, 1988.

\refis{norm1} M.R. Norman, \prl 59, 232, 1987.

\refis{nprl} M.R. Norman, to be published, Phys. Rev. Lett.

\refis{and3} P.W. Anderson, \prb 30, 4000, 1984; K. Ueda and T.M. Rice,
\prb 31, 7114, 1985.

\refis{and2} P.W. Anderson, \prb 32, 499, 1985.

\refis{and1} P.W. Anderson, \prb 30, 1549, 1984.

\refis{ab} P.W. Anderson and W.F. Brinkman, \prl 30, 1108, 1973; in
The Physics of Liquid
and Solid Helium, Part II, ed. K.H. Bennemann and J.B. Ketterson
(J. Wiley, New York, 1978), p. 177.

\refis{naka} S. Nakajima, Prog. Theor. Phys. {\bf 50}, 1101 (1973).

\refis{lvr} K. Levin and O.T. Valls, Phys. Rep. {\bf 98}, 1 (1983).

\refis{ah} J. Appel and P. Hertel, \prb 35, 155, 1987.

\refis{norm} M.R. Norman, Physica C {\bf 194}, 203 (1992).

\refis{js} J.A. Sauls, to be published, Adv. Phys.

\refis{gt} G. E. Volovik and L. P. Gor'kov, JETP {\bf 61}, 843 (1985);
E.I. Blount, \prb 32, 2935, 1985.

\refis{cs} C.H. Choi and J.A. Sauls, \prl 66, 484, 1991; \prb 48, 13684, 1993.

\refis{nabc} M.R. Norman, R.C. Albers, A.M. Boring, and N.E. Christensen,
Sol. State Comm. {\bf 68}, 245 (1988).

\refis{osb} R. Osborn, K.A. McEwen, E.A. Goremychkin, A.D. Taylor, Physica
B {\bf 163}, 37 (1990).

\refis{varma} C.M. Varma, \prl 55, 2723, 1985.

\refis{gold} Z.B. Goldschmidt, \pra 27, 740, 1983.

\refis{miyake} K. Miyake, S. Schmitt-Rink, and C.M. Varma, \prb 34, 6554,
1986; D.J. Scalapino, E. Loh Jr., and J.E. Hirsch, \prb 34, 8190, 1986;
M.T. Beal-Monod, C. Bourbonnais, and V.J. Emery, \prb 34, 7716, 1986.

\refis{pj} W. Putikka and R. Joynt, \prb 37, 2372, 1988; \prb 39, 701,
1989.

\refis{ruln} R.J. Radtke, S. Ullah, K. Levin, and M.R. Norman, \prb  46,
11975, 1992.

\refis{cox2} D.L. Cox, \prl 59, 1240, 1987.

\refis{cox1} D.L. Cox, unpublished.

\refis{coxpc} J.-S. Kang et al, \prb 41, 6610, 1990; D.L. Cox,
private communication.

\refis{aeppli} G. Aeppli et al, \prl 58, 808, 1987; A.I. Goldman et al,
\prb 36, 8523, 1987; P. Frings et al, Physica B {\bf 151}, 499 (1988);
G. Aeppli et al, \prl 60, 615, 1988; G. Aeppli et al, \prl 63, 676, 1989.

\refis{joynt} R. Joynt, J. Magn. Magn. Matls. {\bf 108}, 31 (1992).

\refis{joynt2} R. Joynt, V.P. Mineev, G.E. Volovik, and M.E. Zhitomirsky,
\prb 42, 2014, 1990.

\refis{visser} A. deVisser, A. Menovsky, and J.J.M. Franse, Physica B 
{\bf 147}, 81 (1987).

\refis{steglich} A. Grauel et al, J. Appl. Phys. {\bf 73}, 5421 (1993).

\refis{prni5} K. Andres, S. Darack, and H.R. Ott, \prb 19, 5475, 1979.

\refis{buyers} W.J.L. Buyers and T.M. Holden, in Handbook on the Physics
and Chemistry of the Actinides, Vol. 2, eds. A.J. Freeman and G.H. Lander
(North Holland, Amsterdam, 1985), p. 239.

\refis{gold2} Z.P. Goldschmidt, in Handbook on the Physics
and Chemistry of Rare Earths, Vol. 1, eds. K.A. Gschneidner, Jr. and L.
Eyring (North Holland, Amsterdam, 1978), p. 1.

\refis{free} A.J. Freeman, B.I. Min, and M.R. Norman, in Handbook on the Physics
and Chemistry of Rare Earths, Vol. 10, eds. K.A. Gschneidner, Jr., L.
Eyring, and S. Hufner (North Holland, Amsterdam, 1987), p. 165.

\refis{co} E.U. Condon and H. Odabasi, Atomic Structure (Cambridge Univ.
Pr., Cambridge, 1980).

\refis{judd} B.R. Judd, Operator Techniques in Atomic Spectroscopy
(McGraw-Hill, New York, 1963).

\refis{shalit} A. de-Shalit and I. Talmi, Nuclear Shell Theory (Academic Pr.,
New York, 1963).

\refis{old} K.A. Brueckner, T. Soda, P.W. Anderson, and P. Morel, \pr 118,
1442, 1960; V.J. Emery and A.M. Sessler, \pr 119, 43, 1960.

\refis{bs} N.F. Berk and J.R. Schrieffer, \prl 17, 433, 1966.

\refis{layzer} A. Layzer and D. Fay, Intl. J. Magn. {\bf 1}, 135 (1971).

\refis{lv2} K. Levin and O.T. Valls, \prb 20, 105, 1979; \prb 20, 120, 1979.

\refis{aziz} D.E. Beck, Mol. Phys. {\bf 14}, 311 (1968).

\refis{yip} S. Yip and A. Garg, \prb 48, 3304, 1993.

\refis{su} M. Sigrist and K. Ueda, \revmp 63, 239, 1991.

\refis{leggett} A.J. Leggett, \revmp 47, 331, 1975.

\refis{vv} J.H. VanVleck, \pr 45, 412, 1934.

\refis{racah} G. Racah, \pr 76, 1352, 1949.

\refis{flowers} B.H. Flowers, \prsl A212, 248, 1952;
A.R. Edmonds and B.H. Flowers, \prsl A214, 515, 1952.

\refis{wf} In the LMTO formalism employed here, there are radial functions
and their energy derivatives involved in the variational wavefunctions.
The latter terms are small in the current case and can be ignored in the
coupling constant calculations.

\refis{bw} R. Balian and N.R. Werthamer, \pr 131, 1553, 1963.

\refis{g5} $\Gamma_5$ from Table 5 involves two coefficients ($\alpha$ and
$\beta$) which can be determined by a simple variational method applied
to Eq. 20.  Note they are real numbers.

\refis{stewart} G.R. Stewart, \revmp 56, 755, 1984.

\refis{lee} P.A. Lee, T.M. Rice, J.W. Serene, L.J. Sham, and J.W. Wilkins,
Comm. Cond. Mat. Phys. {\bf 12}, 99 (1986).

\refis{nof} M.R. Norman, T. Oguchi, and A.J. Freeman, \prb 38, 11193, 1988.

\refis{lonz} L. Taillefer and G.G. Lonzarich, \prl 60, 1570, 1988.

\refis{vin} V. Vinokur, J.A. Sauls, and M.R. Norman, unpublished.

\refis{zw} G. Zwicknagl, J. Magn. Magn. Matls. {\bf 76-77}, 16 (1988).

\refis{hayden} S.M. Hayden, L. Taillefer, C. Vettier, and J. Flouquet,
\prb 46, 8675, 1992.

\refis{hess} D.W. Hess, T.A. Tokuyasu, and J.A. Sauls, J. Cond. Mat. Phys.
{\bf 1}, 8135 (1989).

\refis{mach} T. Ohmi and K. Machida, \prl 71, 625, 1993; K. Machida, T. Ohmi,
and M. Ozaki, \jpsj 62, 3216, 1993.

\refis{garg} D.-C. Chen and A. Garg, \prl 70, 1689, 1993; A. Garg and
D.-C. Chen, \prb 49, 479, 1994.

\refis{brison} J.P. Brison, private communication.

\refis{abc} R.C. Albers, A.M. Boring, and N.E. Christensen, \prb 33, 8116, 1986.

\refis{ff} C. Stassis et al, \prb 34, 4382, 1986.

\refis{herbst} J.F. Herbst and R.E. Watson, \prl 34, 1395, 1975.

\refis{cole} P. Coleman, E. Miranda, and A. Tsvelik, \prl 70, 2960, 1993.

\refis{joynt3} R. Joynt, \prl 71, 3015, 1993.

\refis{zl} M.E. Zhitomirski and I.A. Luk'yanchuk, JETP Lett. {\bf 58}, 131
(1993).

\refis{pc} G. Goll, H.v. Lohneysen, I.K. Yanson, and L. Taillefer, \prl 70,
2008, 1993.

\refis{pene} C. Broholm et al, \prl 65, 2062, 1990.

\refis{tu} B.S. Shivaram, Y.H. Jeong, T.F. Rosenbaum, and D.G. Hinks, \prl
56, 1078, 1986.

\refis{tc} K. Behnia et al, \jlowt 84, 261, 1991.

\refis{sh} R.A. Fisher et al, \prl 62, 1411, 1989; K. Hasselbach, L. Taillefer,
and J. Flouquet, \prl 63, 93, 1989.

\refis{hc2} B.S. Shivaram, T.F. Rosenbaum, and D.G. Hinks, \prl 57, 1259, 1986.

\refis{nmr} Y. Kohori et al, \jpsj 57, 395, 1988.

\refis{knight} Y. Kohori et al, \jpsj 56, 2263, 1987.

\refis{n1} The relativistic J=4 pair state is primarily L=5, S=1 in nature,
but this has no relation to the pseudo-spin character of the pair state.

\refis{vd} C.H.H. VanDeurzen, K. Rajnak, and J.G. Conway, J. Opt. Soc. Am. B
{\bf 1}, 45, (1984).

\endreferences

\vfill\eject

\noindent Table 1.  List of known heavy fermion superconductors with the
number of f atoms per unit cell.  In parenthesis is the nature of the
low temperature distorted phase in the single f atom case
(QP - quadrupolar, AF - antiferromagnetic, ? - not fully determined) and
the resulting number of f atoms.

\settabs 2 \columns
\vskip12pt
\hrule
\vskip6pt
\+Case&f Atoms\cr
\vskip6pt
\hrule
\vskip6pt
\+$UPt_3$&2\cr
\+$UBe_{13}$&2\cr
\+$U_2PtC_2$&2\cr
\+$URu_2Si_2$&1 (QP/AF - 2)\cr
\+$UPd_2Al_3$&1 (AF - 2)\cr
\+$UNi_2Al_3$&1 (AF - 2)\cr
\+$CeCu_2Si_2$&1 (? - 2 ?)\cr
\vskip6pt
\hrule

\vfill\eject

\noindent Table 2.  $f^2$ energies in the LS scheme.\refto{racah}

\settabs 2 \columns
\vskip12pt
\hrule
\vskip6pt
\+Term&Energy\cr
\vskip6pt
\hrule
\vskip6pt
\+$^3H$&$E_0 - 9E_3$\cr
\+$^3F$&$E_0$\cr
\+$^3P$&$E_0 + 33E_3$\cr
\+$^1I$&$E_0 + 2E_1 + 70E_2 + 7E_3$\cr
\+$^1G$&$E_0 + 2E_1 - 260E_2 - 4E_3$\cr
\+$^1D$&$E_0 + 2E_1 + 286E_2 - 11E_3$\cr
\+$^1S$&$E_0 + 9E_1$\cr
\vskip6pt
\hrule

\vfill\eject

\noindent Table 3.  $f^2$ energies in the JJ scheme.\refto{flowers}  Note
that the $E_k$ parameters are different from those defined in the LS scheme.

\settabs 2 \columns
\vskip12pt
\hrule
\vskip6pt
\+Term&Energy\cr
\vskip6pt
\hrule
\vskip6pt
\+J = 4&$E_0 - 5E_2$\cr
\+J = 2&$E_0 + 9E_2$\cr
\+J = 0&$E_0 + 3E_1$\cr
\vskip6pt
\hrule

\vfill\eject

\noindent Table 4.  Summary of anisotropic $\Gamma$ for J=4 using parameters
of Figure 6.  n is the type
of bubble used (defined in Eq. 14), $\chi_0$ is the value ($eV^{-1}$) at
which the divergence occurs, and M signifies which $M_J$ state is the
most attractive (with other attractive $M_J$ states listed in parenthesis).
Note there is no attraction for n=5.  For J=2, attraction is found for
$M_J$ = 0 in the n=3,4 cases, but weaker than J=4.  For J=0, no attraction
is found.

\settabs 3 \columns
\vskip12pt
\hrule
\vskip6pt
\+n&$\chi_0$&M\cr
\vskip6pt
\hrule
\vskip6pt
\+0&0.365&4 (3)\cr
\+1&0.365&0 (1,2)\cr
\+2&0.420&0 (2)\cr
\+3&0.420&1\cr
\+4&0.420&0\cr
\+5&0.420&-\cr
\vskip6pt
\hrule

\vfill\eject

\noindent Table 5.  Hexagonal basis functions for J=4.  The forms listed in
this table should be (a) antisymmetrized ($|\mu>|\nu>-|\nu>|\mu>$) and
(b) symmetrized (+
representation) or antisymmetrized (- representation) with respect to
site before use.  For $\Gamma_5$, $\alpha$ and $\beta$ are variational
coefficients such that the sum of their squares is equal to one, and
this representation occurs twice ($\alpha,\beta$ and $\beta,-\alpha$).  Note
that $\Gamma_5$ and $\Gamma_6$ are doublets obtained by replacing
$|\mu>$ by $|-\mu>$.

\settabs 4 \columns
\vskip12pt
\hrule
\vskip6pt
\+Rep&Basis Function\cr
\vskip6pt
\hrule
\vskip6pt
\+$\Gamma_5$&$\alpha |5/2>|3/2> + \beta (0.8018 |5/2>|-1/2> + 0.5976
|3/2>|1/2>)$\cr
\+$\Gamma_3$&$0.7071 |5/2>|1/2> + 0.7071 |-5/2>|-1/2>$\cr
\+$\Gamma_4$&$0.7071 |5/2>|1/2> - 0.7071 |-5/2>|-1/2>$\cr
\+$\Gamma_6$&$0.5345 |5/2>|-3/2> + 0.8452 |3/2>|-1/2>$\cr
\+$\Gamma_1$&$0.2673 (|5/2>|-5/2> +3 |3/2>|-3/2> +2 |1/2>|-1/2>)$\cr
\vskip6pt
\hrule

\vfill\eject

\noindent Table 6.  Coupling constants for $UPt_3$.  These should be
multiplied by the quantity $N\Gamma_4Z^2$ to convert to real
coupling constants.

\settabs 7 \columns
\vskip12pt
\hrule
\vskip6pt
\+Rep&J=4 (+)&J=4 (-)&J=2 (+)&J=2 (-)&J=0 (+)&J=0 (-)\cr
\vskip6pt
\hrule
\vskip6pt
\+$\Gamma_5$&0.069&0.073\cr
\+$\Gamma_5$&0.029&0.101&0.049&0.071\cr
\+$\Gamma_3$&0.024&0.064\cr
\+$\Gamma_4$&0.013&0.120\cr
\+$\Gamma_6$&0.018&0.125&0.035&0.106\cr
\+$\Gamma_1$&0.076&0.114&0.065&0.099&0.495&0.057\cr
\vskip6pt
\hrule

\vfill\eject

\figurecaptions

\figis{1} Interaction potential between two He atoms.\refto{aziz}  The
interaction potential of two f electrons of a uranium ion would
look similar with appropriately scaled axes (with attraction in that
case due to the ion core).

\figis{2} $f^2$ multiplet structure of a $U^{4+}$ ion.\refto{gold,vd}
Energy is plotted versus J, with labels referring to L.

\figis{3} Effective interaction of Eq. 10 versus $\chi_0$.

\figis{4} Effective interaction of Eq. 11 (triplet, lower curve) and
Eq. 12 (singlet, upper curve) versus $\chi_0$.  This is very similar to
the effective interaction in the $^3He$ problem.

\figis{5} Effective interaction (LS) in eV for $^3H$, $^3F$, $^3P$, and $^1I$
versus $\chi_0$ for parameters appropriate to a U
ion\refto{gold,herbst} ($E_0$ = 1225 meV,
$E_1$ = 470.3 meV, $E_2$ = 1.923 meV, $E_3$ = 43.28 meV).  The zeros of
energy for the $f^1$ and $f^2$ cases are marked by the dashed lines.

\figis{6} Effective interaction (JJ) in eV for J=4,2,0
versus $\chi_0$ for a U ion.  Same parameters and notation as in Figure 5.
 
\figis{7} Effective interaction (JJ) in eV for J=4,2,0
versus $\chi_0$ for parameters appropriate to a Ce
ion\refto{gold2,free} ($E_0$ = 6000 meV,
$E_1$ = 484.5 meV, $E_2$ = 2.293 meV, $E_3$ = 47.67 meV; note these are
LS $E_k$).
The zero of energy is the dashed line.

\figis{8} Plot of the k points used in the calculations on $UPt_3$ in
symmetry planes of the Brillouin zone constructed from a uniform grid
within an energy cutoff of 1 mRy of the Fermi energy.  Lines represent
the Fermi surface.

\figis{9} Plots of $|\Delta_{\sigma}(\vec k)|$ with $\sigma$ = + for
$\Gamma_6^-$
state on the grid of points of Figure 8.
Size of the dots represent the magnitude of the gap.
Where no dots appear, gap is zero or very small.

\figis{10} Same as Figure 9, but with $\sigma$ = -.

\figis{11} Smoothed fit to the density of states (normalized to the
normal state value) constructed from
the gaps plotted in Figures 9 and 10.  Energy units are normalized
to the value of the largest gap.

\endfigurecaptions

\vfill\eject

\endit

FIGURES BEGIN HERE

This file needs to be stripped off and the following done.  Type uudecode (file)
(resulting file will be called fig.tar.Z).  Type uncompress fig.tar.Z
Resulting file will be fig.tar  Type tar -xvf fig.tar  11 postscript figures
called fig#.ps will appear
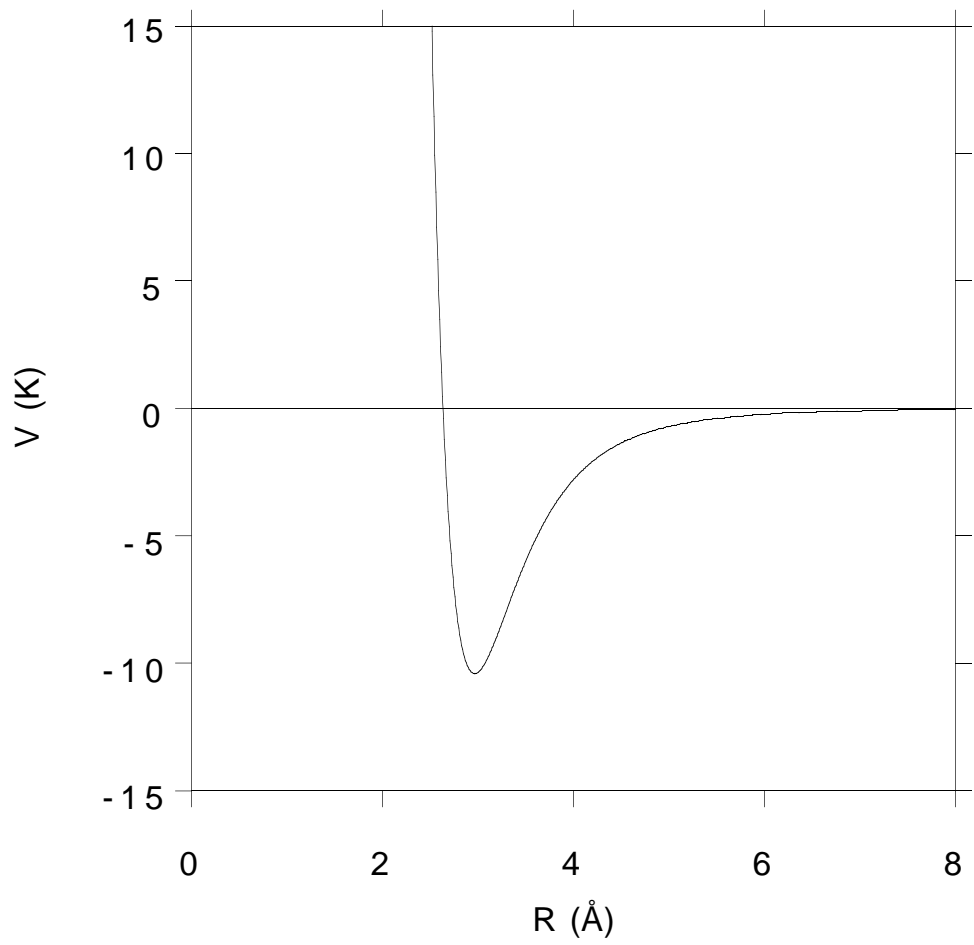
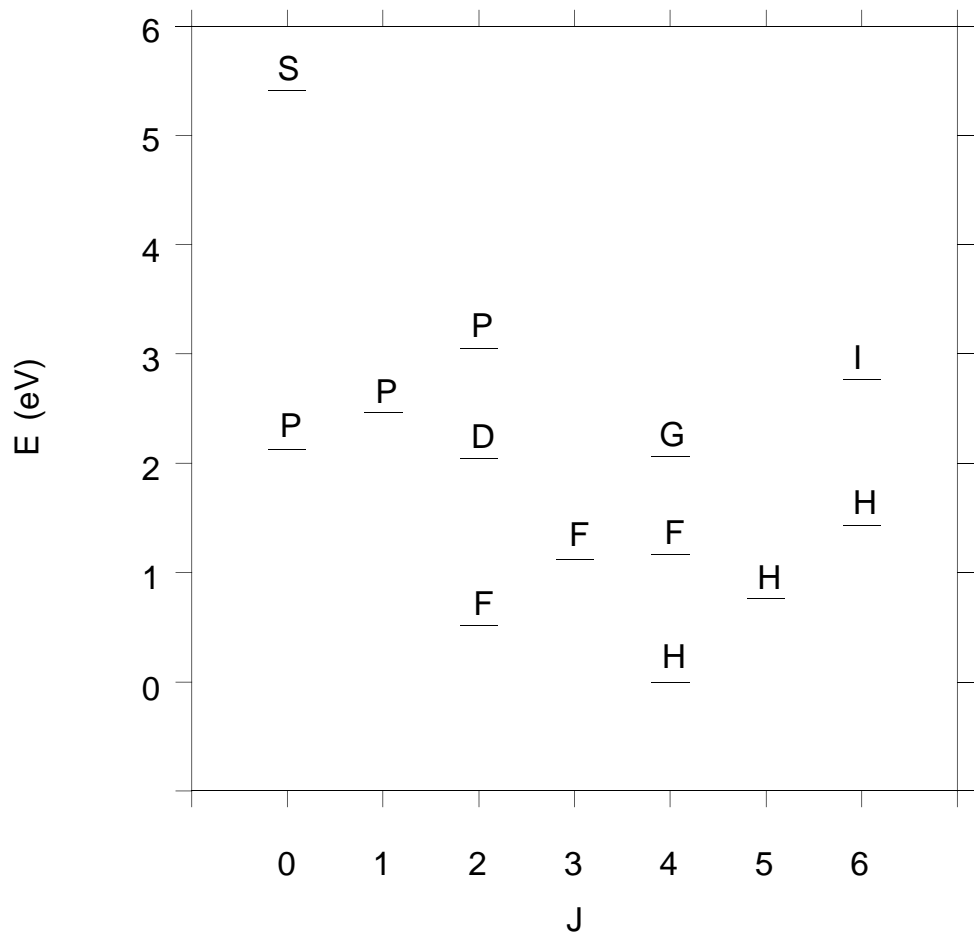
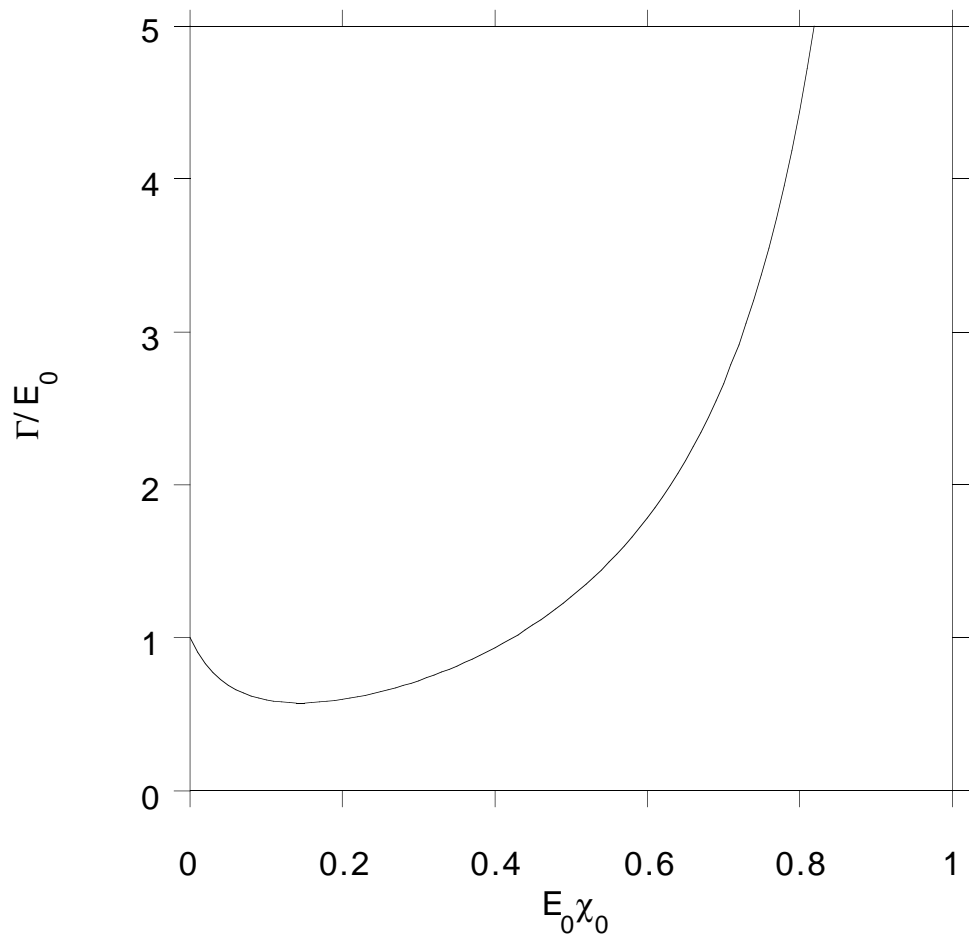
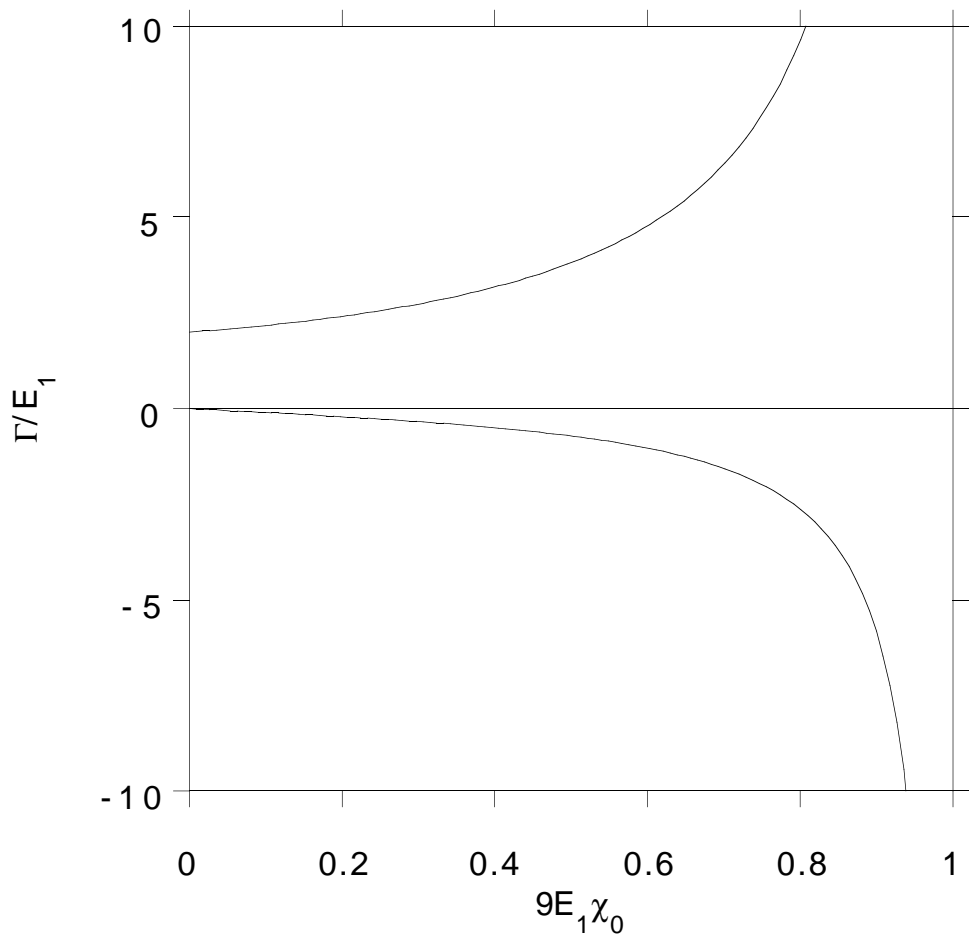
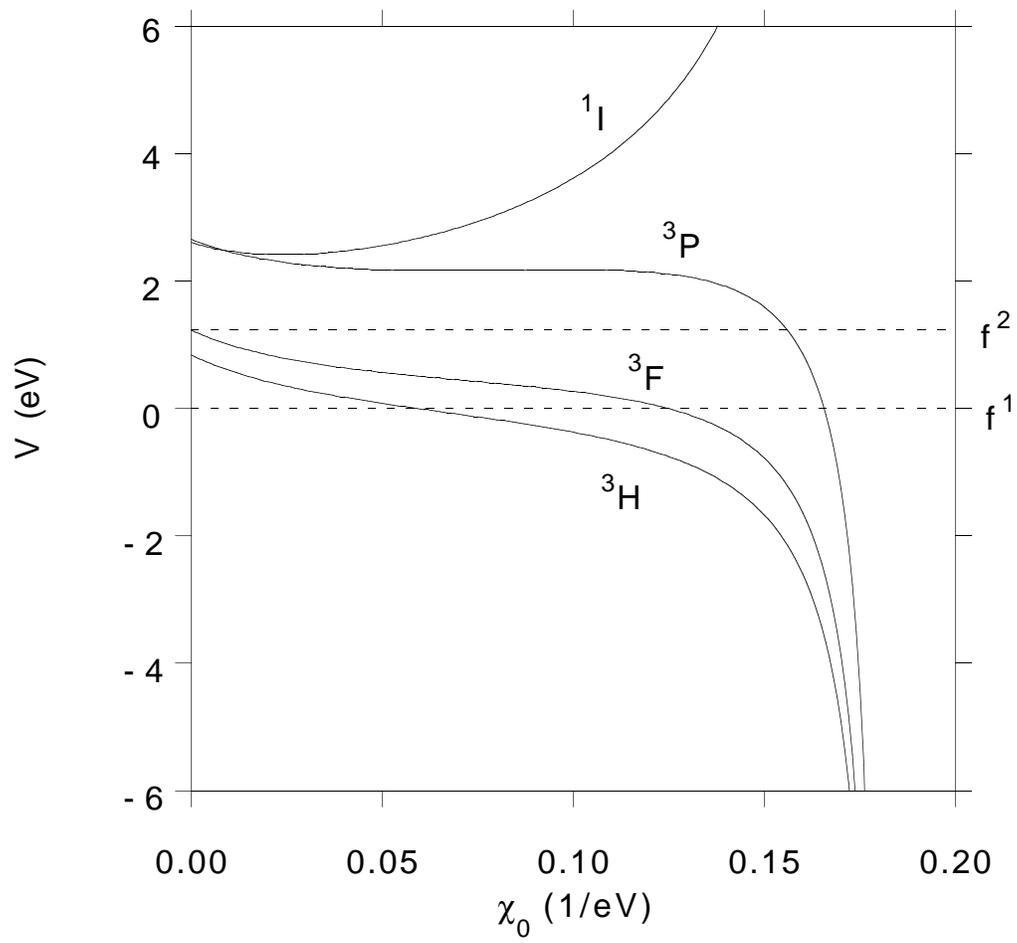
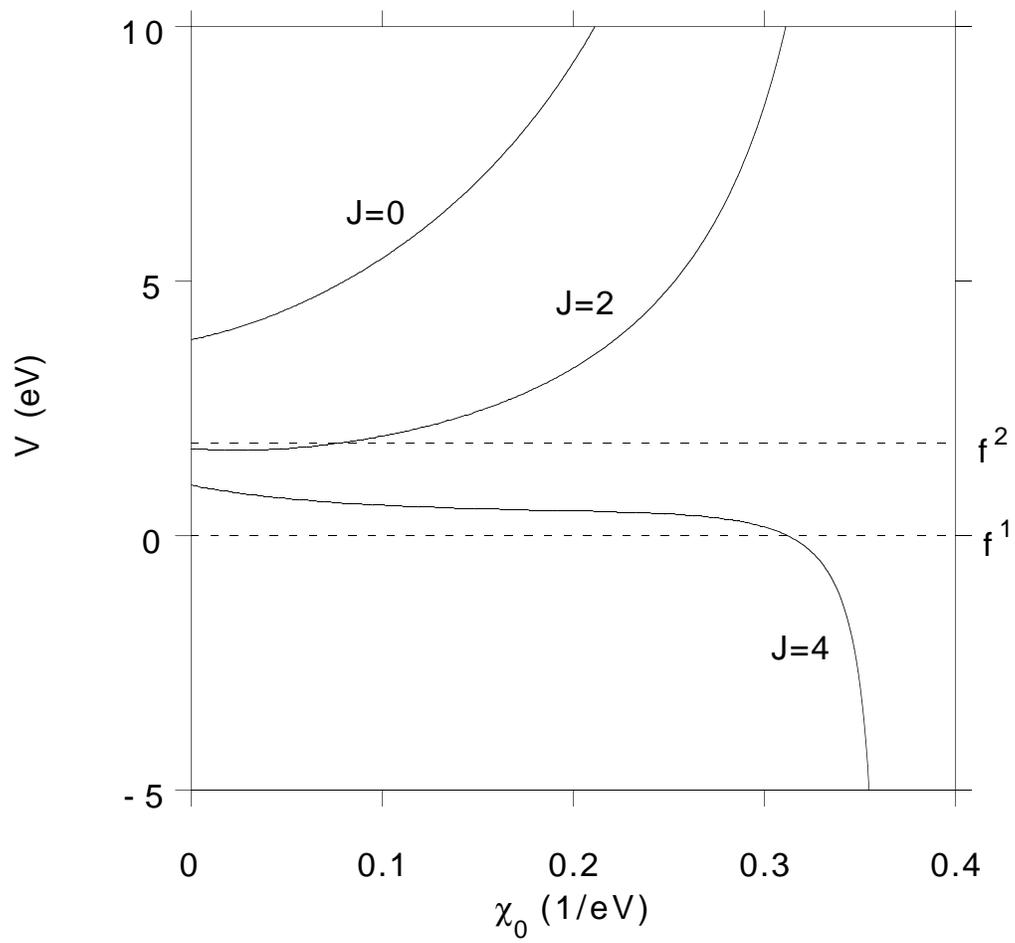
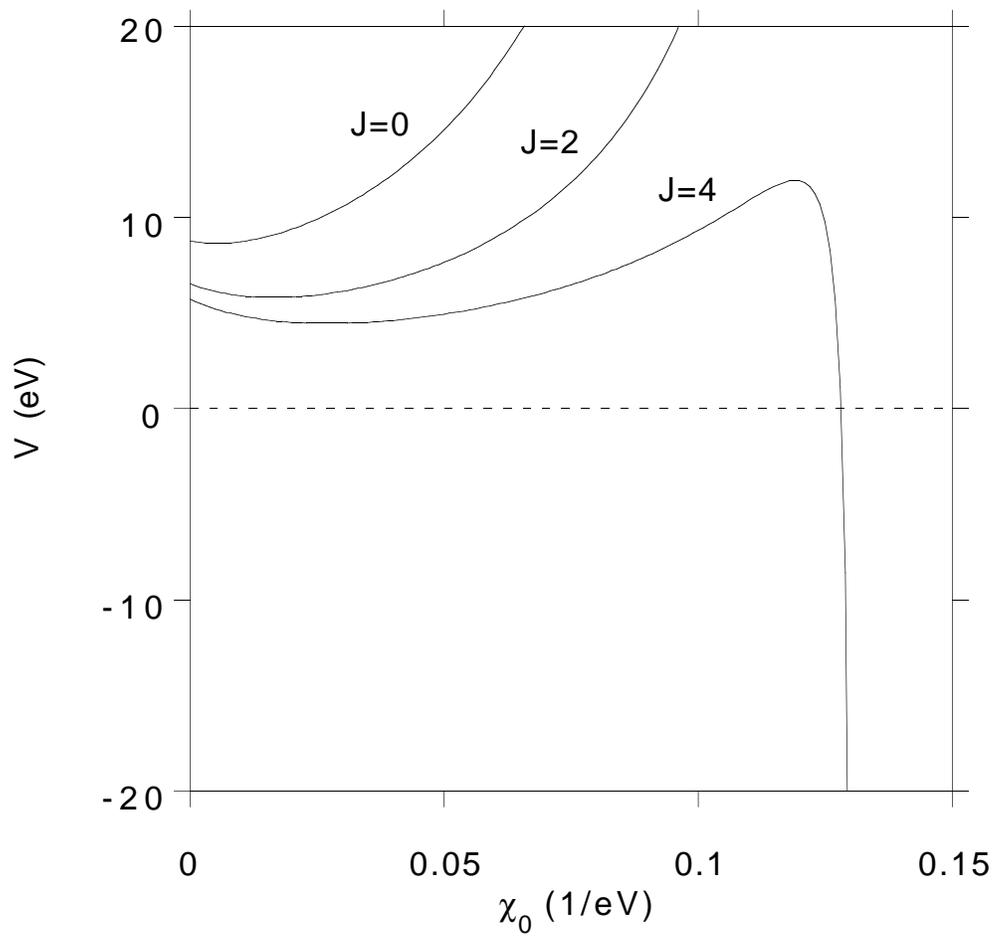
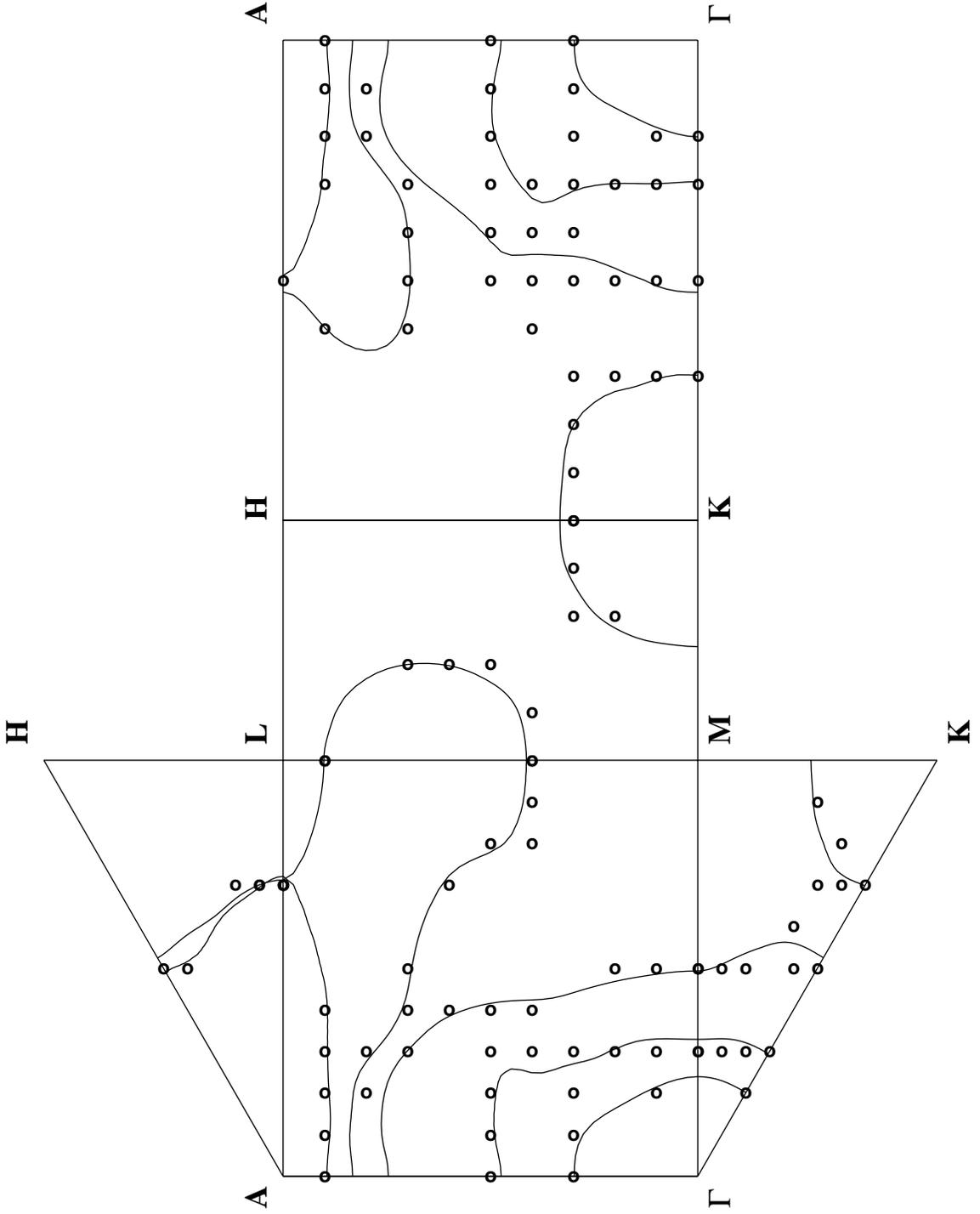
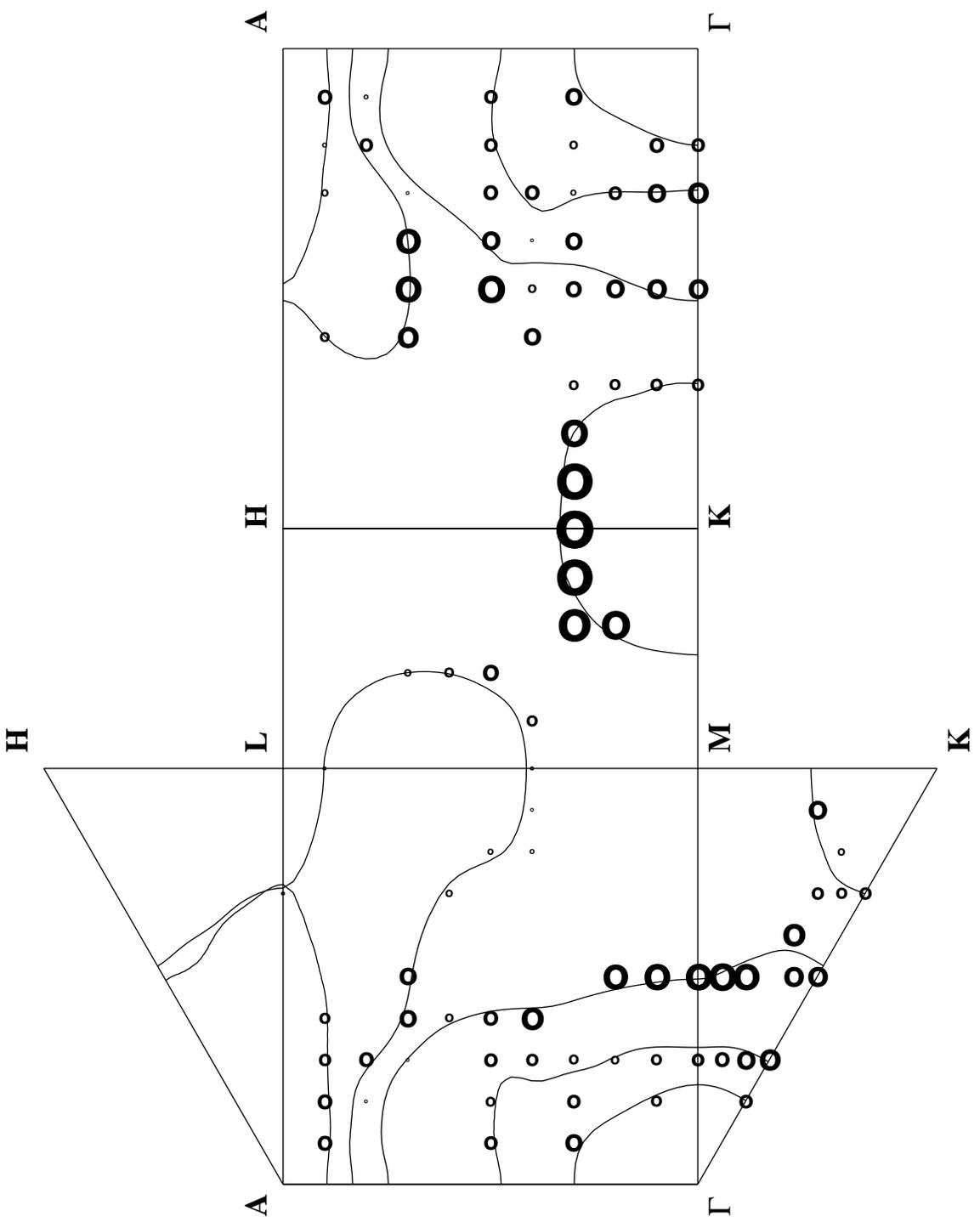
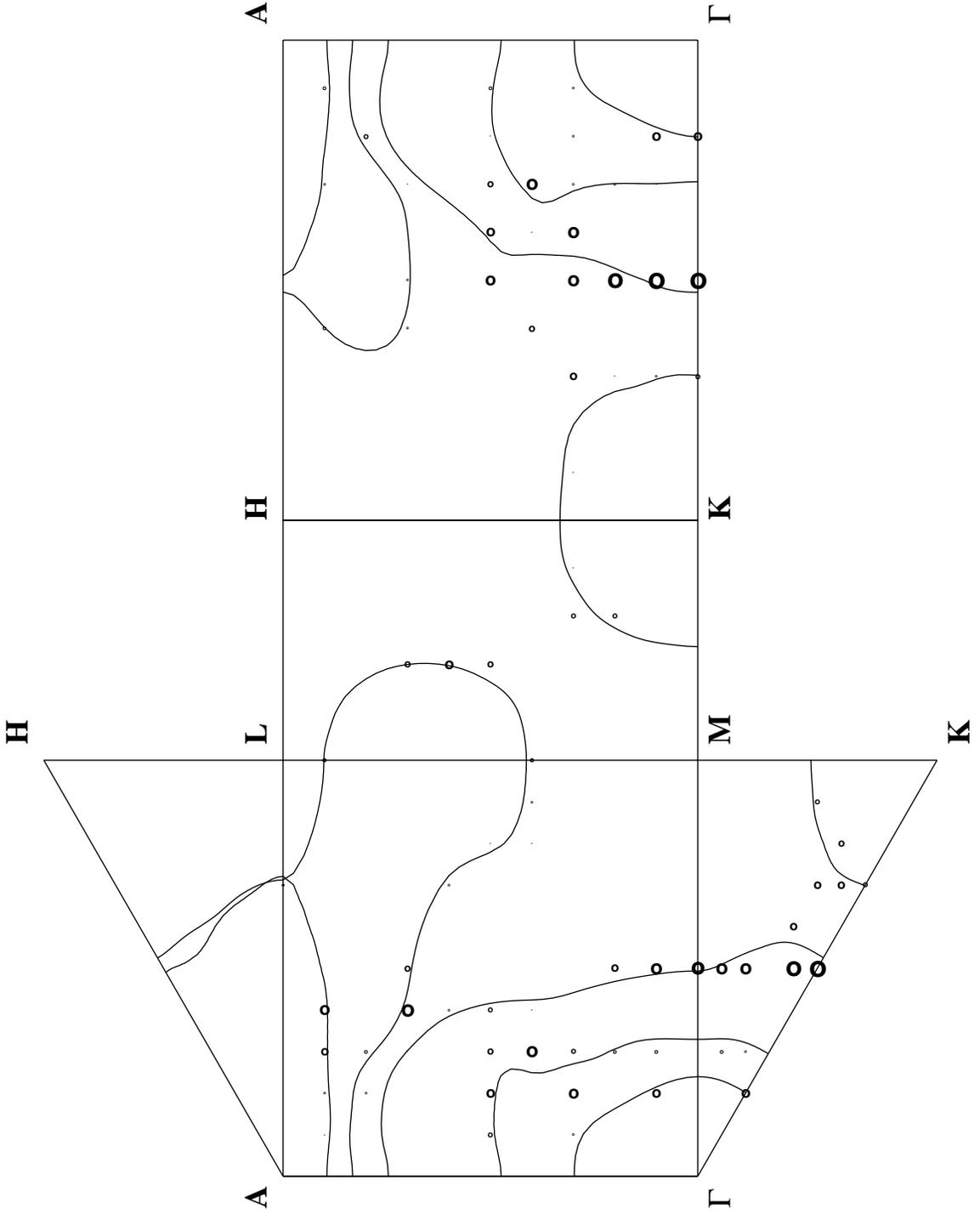
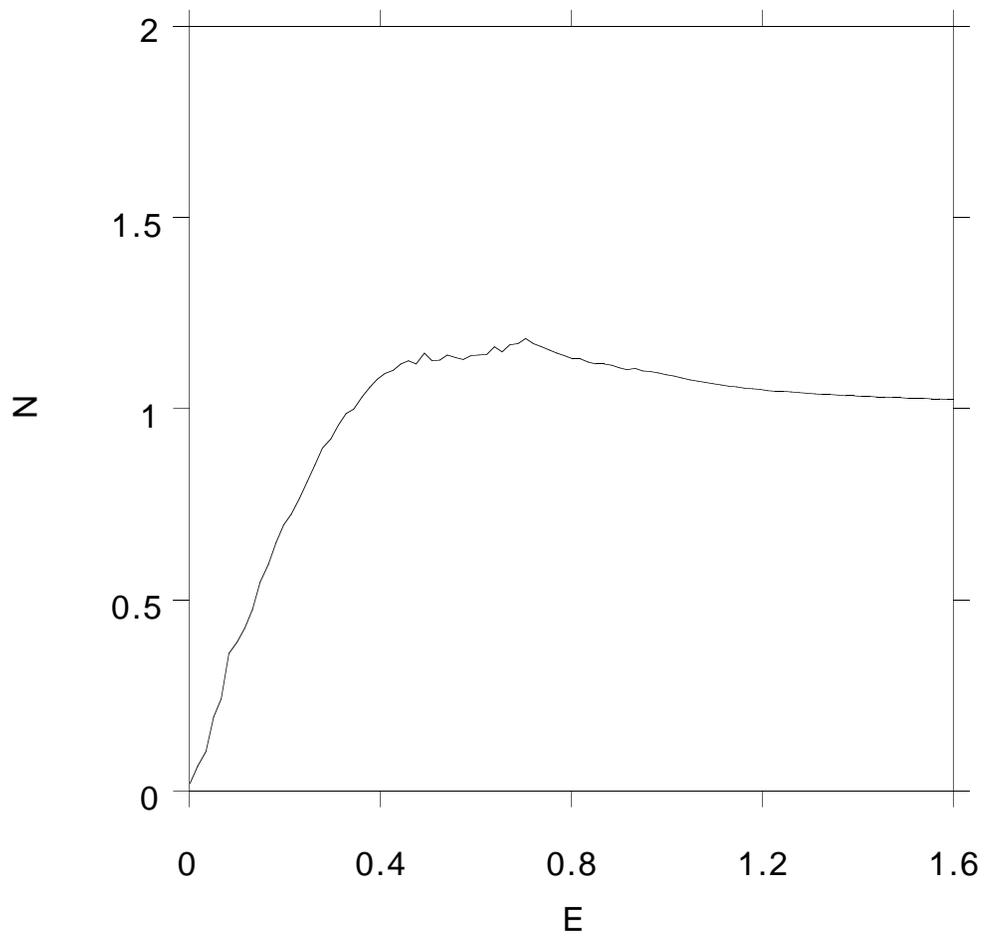
